\documentclass[conference,compsoc]{IEEEtran}
\IEEEoverridecommandlockouts
\usepackage{epsfig,endnotes}
\usepackage{mdframed}
\usepackage{graphicx}
\usepackage{algorithm}
\usepackage{listings}
\usepackage{gensymb}
\usepackage[utf8]{inputenc}
\DeclareUnicodeCharacter{2212}{-}
\usepackage{mathtools}
\usepackage{algpseudocode}
\usepackage{multirow}
\usepackage{amssymb}
\usepackage{url}
\usepackage{hyperref}
\usepackage{xcolor,colortbl}
\usepackage{caption}    
\usepackage{array}
\usepackage{booktabs}
\usepackage{pifont}
\usepackage{verbatim}
\usepackage{tikz}

\usepackage{makecell}

\usepackage{amsmath, amssymb, amsfonts}
\usepackage{enumitem}
\usepackage{siunitx}

\usepackage{listings}
\usepackage{xcolor}
\definecolor{green}{rgb}{0.01, 0.75, 0.24}
\definecolor{red}{rgb}{0.76, 0.23, 0.13}
\usepackage{pifont}
\newcommand{\cmark}{\textcolor{green}{\ding{51}}}%
\newcommand{\xmark}{\textcolor{red}{\ding{55}}}%

\renewcommand{\paragraph}[1]{\medskip \noindent {\bf #1.}}

\newcommand{\lstbg}[3][0pt]{{\fboxsep#1\colorbox{#2}{\strut #3}}}
\lstdefinelanguage{diff}{
  basicstyle=\ttfamily\small,
  morecomment=[f][\lstbg{red!20}]-,
  morecomment=[f][\lstbg{green!20}]+,
  morecomment=[f][\textit]{@@},
}

\definecolor{mGreen}{rgb}{0,0.58,0}
\definecolor{mGray}{rgb}{0.5,0.5,0.5}
\definecolor{mPurple}{rgb}{0.58,0,0.82}
\definecolor{backgroundColour}{rgb}{0.80,0.80,0.80}

\lstdefinelanguage{markdown}{
  morekeywords={},
  sensitive=false, 
  basicstyle=\footnotesize\ttfamily,
  columns=fullflexible,
  breaklines=true,      
  frame=single, 
  morecomment=[l]{\#},
  morestring=[b]",
}

\lstdefinestyle{CStyle2}{
  backgroundcolor=\color{white},
  basicstyle=\footnotesize\ttfamily,
  columns=fullflexible,
  breakatwhitespace=false,      
  breaklines=true,                
  captionpos=b,                    
  commentstyle=\color{mGreen}, 
  extendedchars=true,              
  frame=single,                   
  keepspaces=true,             
  keywordstyle=\color{blue},      
  language=c++,                 
  numbers=left,                
  numbersep=5pt,                   
  numberstyle=\tiny\color{mGray}, 
  rulecolor=\color{mGray},        
  showspaces=false,               
  showtabs=false,                 
  stepnumber=1,                  
  stringstyle=\color{magenta},    
  tabsize=1,                      
  title=\lstname,
  morecomment=[f][\lstbg{red!20}]-,
  morecomment=[f][\lstbg{green!20}]+,
  morecomment=[f][\textit]{@@},
}

\definecolor{mGreen}{rgb}{0,0.58,0}
\definecolor{mGray}{rgb}{0.5,0.5,0.5}
\definecolor{mPurple}{rgb}{0.58,0,0.82}
\definecolor{backgroundColour}{rgb}{0.80,0.80,0.80}

\begin{document}

\date{}

\title{\Large \bf ZeroLeak: Using LLMs for Scalable and Cost Effective Side-Channel Patching}

\author{
{\rm M. Caner Tol}
\\
Worcester Polytechnic Institute\\
mtol@wpi.edu
\and
{\rm Berk Sunar}\\
Worcester Polytechnic Institute\\
sunar@wpi.edu
}

\maketitle
\thispagestyle{plain}
\pagestyle{plain}

\subsection*{Abstract}
Security critical software, e.g., OpenSSL, comes with numerous side-channel leakages left unpatched due to a lack of resources or experts. The situation will only worsen as the pace of code development accelerates, with developers relying on Large Language Models (LLMs) to automatically generate code. In this work, we explore the use of LLMs in generating patches for vulnerable code with microarchitectural side-channel leakages. For this, we investigate the generative abilities of powerful LLMs by carefully crafting prompts following a zero-shot learning approach. All generated code is dynamically analyzed by leakage detection tools which are capable of pinpointing information leakage at the instruction level leaked either from secret dependent accesses or branches or vulnerable Spectre gadgets, respectively. Carefully crafted prompts are used to generate candidate replacements for vulnerable code which are then analyzed for correctness and for leakage resilience. 

After extensive experimentation, we determined that the way prompts are formed and stacked over a series of queries plays a critical role in the LLMs' ability to generate correct and leakage-free patches. We develop a number of \textit{tricks} to improve the chances of correct and side-channel secure code. Moreover, when we compare various LLMs, we found that OpenAI's GPT4 is far superior compared to Google PaLM and Meta LLaMA in generating patches with nearly all leakages fixed in a microbenchmark of vulnerable codes as well as Spectre v1 gadgets. We also found that  GPT4 is more successful than GPT3.5 in generating both correct and secure code, with many failed attempts observed in the latter. As for efficiency, GPT4 provides a far more efficient patch with up to 10 times less overhead when compared to the clang compiler-supported lfence Spectre mitigation. 
From a cost/performance perspective, the GPT4-based configuration costs in API calls a mere few cents per vulnerability fixed. We note, however, that there is great variability in cost, depending on the type of vulnerability (leak vs. Spectre gadget) and length of the vulnerable code patched. Regardless our results show that LLM-based patching is far more cost-effective and thus provides a scalable solution. Finally, the framework we propose will improve in time, especially as vulnerability detection tools and LLMs mature. 

\section{Introduction}
The advent of microarchitectural attacks has instigated efforts to mitigate vulnerabilities in hardware/firmware and in deployed software libraries. Earlier vulnerabilities, such as those exploiting secret dependent execution time and cache/memory access patterns, were followed by more advanced attacks exploiting microarchitectural optimizations such as out-of-order and speculative execution~\cite{Lipp2018meltdown,kocher2019spectre}, transient write forwarding and shared buffers~\cite{ridl,canella2019fallout,Schwarz2019ZombieLoad}. 

One of the earliest and still most accessible forms of side-channel leakage is execution time. If a developer inadvertently writes code, e.g., with secret data-dependent branches, by measuring the execution time, an attacker can deduce secret information. Therefore, identifying vulnerable software and replacing them with their constant-time versions has been a goal of security researchers. This is challenging in practice since repositories have complex interdependence with many potentially vulnerable pieces, while their execution time is also dependent on many factors, e.g., the platform and its configuration, the compiler. 

Spectre was first discovered and publicly disclosed by security researchers in the original Spectre paper in 2018~\cite{kocher2019spectre}. Spectre v1 occurs when attackers can trick the CPU into speculatively executing code that would not normally be run during normal program execution. By exploiting this vulnerability, attackers can potentially access sensitive data or information stored in the memory of other applications or the operating system. The attack leverages the processor's speculative execution to infer this sensitive data and exfiltrate it. 

In his blog, Kocher \cite{kocher2018spectre} shared 15 code snippets vulnerable to variations of Spectre v1 (Spectre gadgets) to test out a new version of Microsoft VC/C++ compiler with built-in mitigation~\cite{msvcspectre} based on the addition of the \texttt{LFENCE} instruction to sensitive parts of the code identified by Microsoft's static analyzer. The compiler only manages to mitigate Spectre in the first two gadgets. Kocher points out that his code samples are far from comprehensive, e.g., they all rely on cache modification as a covert channel, and they all reside in simple functions more amenable to static analysis. Cauligi et al.\cite{cauligi2021sok} present a comprehensive survey of existing Spectre v1 defenses and non-constant time detection tools e.g. \texttt{oo7}~\cite{wang2019oo7}, Spectector~\cite{guarnieri2020spectector}, SpecFuzz~\cite{OleksenkoSpecFuzz18}, Pitchfork~\cite{cauligi2020constant}.

Code with microarchitectural vulnerabilities, e.g., secret dependent non-constant time or code vulnerable to Spectre v1 has since been a significant concern for the tech industry. Hardware and software vendors have released mitigations to reduce the risk of exploitation, but fully addressing these vulnerabilities remains an ongoing challenge. Moreover, these mitigations often come at the cost of decreased performance, as they may disable or limit certain speculative execution features.

In a study among the crypto library developers, 61.4\% of the participants stated that they are either not aware or they do not use the tools for testing and verifying the constant-timeness~\cite{jancar2022they} -- a necessary but insufficient condition for side-channel security. To make matters worse, many of these libraries that are used by millions of end-users are managed by a small number of developers in open-source projects. They neither possess the knowledge nor the resources to patch their software against such low-level leakages. Often times reported vulnerabilities go ignored and unpatched in publicly available open-source crypto libraries used by millions, e.g., see Microwalk-CI~\cite{jan2022microwalk}, due to lack of resources. Another striking example is in the OpenSSL Blog Post \cite{OpenSSLBlog} explaining their decision on why they chose \textit{not} to patch for newly discovered Spectre gadgets reported in \cite{MosierAxiomatic22} :
``\textit{Most potentially vulnerable code is extremely non-obvious, even to experienced security programmers. It would thus be quite easy to introduce new attack vectors or fix existing ones unknowingly.}'' and ``\textit{Automated verification and testing of the attacks is necessary but not sufficient. We do not have automated detection for this family of vulnerabilities and if we did, it is likely that variations would escape detection.}''. These comments highlight the need for reliable and transparent patch automation.

In this work, we study the use of LLMs for automated patching of security-critical software. Indeed, it is expected that 80\% of the software development lifecycle will use generative AI, i.e., LLMs, by 2025~\cite{gartner2023}. Thus, evaluating LLMs' capability to generate security-critical implementations is an urgent need.
What happens if we use ordinary prompts to generate crypto code, and how can we improve code generation to improve side-channel security while ensuring functional correctness? We are encouraged by rapid advances in LLMs. Fueled by recent innovations in Transformer networks, generative models, and the availability of massive datasets and large compute clusters, it has become possible to train Large Language Models (LLMs). LLMs such as GPT3~\cite{brown2020language} and GPT4~\cite{openai2023gpt4} by OpenAI, BERT~\cite{devlin2019bert} and PaLM2~\cite{anil2023palm} by Google, RoBERTa~\cite{liu2019roberta} and LLaMA~\cite{touvron2023LLaMA, touvron2023LLaMAtwo} by Meta AI have shown impressive performance in AI applications and in natural language processing (NLP). These tools are also trained using code snippets, allowing one to parse and even generate code in common programming languages flexibly. 

In this work, we study the use of LLMs in a zero-shot approach in concert with state-of-the-art leakage and vulnerability detection tools to fix data-dependent non-constant time behavior, as well as secret-dependent branching and Spectre v1 gadgets. Such vulnerabilities are known to exist in numerous security libraries deployed on millions of machines. Yet, due to the lack of resources, i.e., experts and financial resources, they go unpatched. Our goal is to make use of the massive recent advances in LLMs such as OpenAI GPT, Google PaLM, and Meta LLaMA to generate patches automatically. Note that LLMs are fairly large, and it takes weeks to months to train on massive datasets, resources that only large companies have access to. Our goal is to utilize LLMs via API access to bring down the cost of patch deployment to cents per microarchitectural leakage. 

\subsection{Contributions}
\begin{itemize}[leftmargin=*]
\item We present the first comprehensive study of state-of-the-art LLMs, i.e., OpenAI GPT, Google PaLM 2, and Meta LLaMA, to automatically patch microarchitectural vulnerabilities such as secret dependent (non-constant time) code and Spectre v1 gadgets. 

\item We build a toolchain that tests binaries for leakage and Spectre detection tools, specifically Microwalk~\cite{jan2022microwalk}, Pitchfork~\cite{cauligi2020constant}, Spectector~\cite{guarnieri2020spectector}, and KLEESpectre~\cite{wangKLEESpectre2020}, and then automatically generates security patches to be included in the source files using LLMs.

\item From a Continuous Integration/Continuous Development (CI/CD) perspective in  the software development life cycle, the proposed framework allows us to patch the source code (e.g., C/C++, Javascript, etc.) while testing the binary after compilation on a target machine. Compared to binary patching, we retain the ability to review and revise the source. At the same time, we are also taking into account the effect of the compiler and platform configuration on security and efficiency by testing the binary for leakage. This approach allows us to continuously improve the software as hardware systems and software stacks evolve.

\item 
On a microbenchmark of C code we compiled from known vulnerabilities, GPT4 excels in patching 97\% of all leakages successfully of every type of patching points in the benchmark, while the total cost of patching 33 leaks is at \$1.34. GPT3.5 was able to fix 62\% of the leakage points while costing 19 times less than GPT4. Google \texttt{chat-bison} and Meta LLaMa2 patch 56\% and 35\% across all vulnerabilities, respectively, albeit at much lower cost.

\item Our framework is only limited by the capability of the detection tools, e.g., false positives and negatives, and will rapidly improve further with better detection tools. Similarly, LLMs are improving at an astounding rate (almost every month, a new LLM is released), and we expect significant improvement in the overall performance of our tool. 

\item From an efficiency perspective, with up to $\sim10\times$ faster than Spectre v1 patches generated with existing methods, our toolchain significantly outperforms compiler-based techniques such as in \texttt{clang} \texttt{lfence} injection. Hence, the proposed approach provides an opportunity to remove unnecessary inefficiencies while retaining security.

\item Since we are patching the source code with the output generative LLM, the patches are also commented, which makes it easier to make sense of the patch and maintain the code.

\end{itemize}

\section{Background}
\subsection{Constant-Time Implementations}
Constant-time implementations refer to cryptographic algorithms and methods that take a constant amount of time to execute, regardless of the input size or values. This type of implementation is essential for securing systems against timing attacks, which are a type of side-channel attack where an attacker can gain information about a system's secret data by observing the execution times. A practical example of this could be seen in the RSA decryption algorithm, where different execution paths can be chosen based on the secret key bit. An attacker can utilize this timing discrepancy to infer the secret key~\cite{kocher1996timing}. The implementation process of constant-time cryptographic algorithms typically requires meticulous programming to ensure that no branches (such as if-then-else constructs), loops, or other operations are contingent on the secret data. For instance, cryptographic algorithms like AES should avoid data-dependent branches and employ bit-wise operations, which are known to execute in constant time on most platforms.
\begin{figure}[h]
\footnotesize
\begin{lstlisting}[style=CStyle2,linewidth=0.981\columnwidth,xleftmargin=0.35cm]
bool equals(byte a[], size_t a_len,
            byte b[], size_t b_len) {
  for (size_t i = 0; i < a_len; i++)
    if (a[i] != b[i])  // data dependent!
      return false
  return true;
}
\end{lstlisting}
\vspace{-0.7cm}
\caption{An example of a data-dependent equality check logic that violates the constant-time property. Adapted from~\cite{intel2022guidelines}.}
\label{fig:leaky_snippet}
\end{figure}

Challenges exist in guaranteeing a truly constant-time implementation, particularly on contemporary CPUs that possess features like out-of-order execution and speculative execution. This necessitates an in-depth understanding of both the cryptographic algorithm and the hardware it functions on.

There are several examples of constant-time cryptographic algorithms, such as the \textit{constant-time carry-less multiplication} utilized in AES-GCM implementations and the \textit{constant-time modular inversion} employed in elliptic curve cryptography.

A plethora of tools exist for automated verification of the constant-time criterion. However, there is a significant discrepancy between academic research and cryptographic engineering practice. Despite the availability of tools for checking constant-time execution, developers often overlook this due to resource constraints~\cite{jancar2022they}. 

Considering the escalating sophistication of side-channel attacks, the increasing heterogeneity, and the constant evolution of computing platforms, security-critical software needs to be continuously reevaluated for constant-time execution. Future research and developmental efforts will perpetually focus on generating more secure and efficient constant-time algorithms.

{\bf Microwalk-CI}~\cite{jan2022microwalk} is a dynamic side-channel analysis framework for easy integration into a JavaScript development workflow. Microwalk-CI adapts the existing Microwalk \cite{jan2018microwalk} framework, which was originally designed for finding leakages in binary software. For this, Microwalk generates a number of execution traces for a set of random inputs and then compares them using mutual information (MI), a robust measure that allows quantitatively assess the extent of information leakage. MI can capture a wide range of potential leakages, including those from the execution path and memory accesses. However, it is worth noting that \textit{Microwalk} requires the tester to generate an input template file for each function tested and requires interpretation of the report file as it generates entropy estimates.

\subsection{Spectre v1}
Spectre v1 (CVE-2017-5753), also known as \textit{Bounds Check Bypass}, affects a wide range of modern processors, including those from Intel, AMD, and ARM. It allows an attacker to trick a program into speculatively executing code that should not have been executed, potentially leaking sensitive data. Figure~\ref{lst:spectre} provides a simple Spectre v1 example. Modern CPUs have components for predicting conditional branch outcomes to execute the instructions speculatively. The attacker fills the branch history table with malicious entries such that the branch predictor has a high chance of misprediction when the victim runs the branch. 
\begin{figure}[h]
\footnotesize
\begin{lstlisting}[style=CStyle2,linewidth=0.981\columnwidth,xleftmargin=0.35cm,language=C]
void victim_function(size_t x){
	if(x < size)
		temp &= array2[array1[x] * 512];
}
\end{lstlisting}
\vspace{-0.7cm}
\caption{Sample Spectre-V1 C Code\label{lst:spectre}}
\end{figure}
The $2^{nd}$ line checks whether the user input \texttt{x} is in the bound of \texttt{array1}. During a regular execution environment, if the condition is satisfied, the program retrieves $x^{th}$ element of \texttt{array1}, and a multiple of the retrieved value (512) is used as an index to access the data in \texttt{array2}. However, under some conditions, the \texttt{size} variable might not be present in the cache. In such occurrences, instead of waiting for \texttt{size} to be available, the CPU executes the next instructions speculatively. To eliminate unnecessary pipeline stalls when \texttt{size} becomes available. Once the CPU notices the misprediction, it rolls back and follows the correct execution path. Although the results of speculatively executed instructions are not observable in architectural components, the access to the \texttt{array2} leaves a footprint in the cache, making it possible to leak the data through side-channel analysis. 

Spectre v1 is challenging to mitigate because it is a hardware-level issue, and traditional software-based security measures are not sufficient to fully protect against it. Since Spectre v1 is a complex vulnerability with widespread implications across different processor architectures and generations, it has been an ongoing challenge for the industry to address comprehensively.

\subsubsection{Testing for Spectre-v1}
We briefly summarize Pitchfork~\cite{cauligi2020constant}, Spectector~\cite{guarnieri2020spectector}, and KLEESpectre~\cite{wangKLEESpectre2020}.


{\bf KLEESpectre}~\cite{wangKLEESpectre2020} aims to model cache usage with symbolic execution to detect speculative interference, which is based on KLEE symbolic execution engine. KLEESpectre is evaluated on Kocher's Spectre v1 variants \cite{kocher2018spectre} and on cryptographic
libraries \texttt{libTomCrypt}, \texttt{Linux-tegra},
\texttt{openssl} and \texttt{hpn-ssh}. 

{\bf Pitchfork}~\cite{cauligi2020constant} is a symbolic analysis tool that verifies that code is constant-time with respect to secret values such as encryption keys or message plaintexts. Pitchfork uses underconstrained symbolic execution augmented with dynamic taint tracking to verify constant-time execution. In particular, it uses a shadow memory to track secrets even as they are stored and loaded from memory. Pitchfork also allows the specification of function hooks.  This allows Pitchfork to verify a code at the protocol level while ignoring the implementation of crypto primitives. Pitchfork was used to verify that a large portion of Mozilla’s NSS cryptographic library is constant-time while also finding several constant-time vulnerabilities.

{\bf Spectector}~\cite{guarnieri2020spectector} introduces the notion of  
{\em speculative non-interference} (SNI), and develops an algorithm based
on symbolic execution to automatically prove SNI or detect violations indicating Spectre vulnerabilities which then can be patched. SNI requires that speculatively executed instructions do not leak more information into the microarchitectural state than what the intended behavior does, i.e., what is leaked by the standard, non-speculative semantics. To capture “leakage into the microarchitectural state”, we consider an observer of the program execution that sees the locations of memory accesses and jump targets. Under this observer model, an adversary may distinguish two initial program states if they yield different traces of memory locations and jump targets. 
Spectector is able to detect all leaks in the 15 Spectre v1 variants by Kocher \cite{kocher2018spectre} and also detect novel, subtle leaks that are out of scope of \texttt{oo7}~\cite{wang2019oo7}, and even identify cases
where compilers unnecessarily inject countermeasures.

\section{Related Work}
The field of automated program repair has seen various advances, but these studies typically focus on syntactic and build errors, with fewer exploring the domain of security vulnerabilities, and none, to date, have addressed the issue of microarchitectural vulnerabilities.

DeepFix, as Gupta et al.~\cite{gupta2017deepfix} proposed, aims to automatically correct common programming errors using a sequence-to-sequence neural network model. However, this method is fundamentally limited in scope, as it does not tackle any security vulnerabilities. Its performance is also contingent on the accuracy of error location prediction, which is inherently challenging.
Similarly, the Break-It-Fix-It (BIFI) method by Yasunaga et al.~\cite{yasunaga2021break} primarily targets syntactic errors, leaving the important domain of security vulnerabilities unaddressed. Moreover, despite improving over previous methods, BIFI's repair accuracy still leaves a significant percentage of errors uncorrected, pointing towards a potential need for better training methods and error diversity.
The Graph2Diff model introduced by Tarlow et al. ~\cite{tarlow2020learning} extends the focus to build errors but continues to overlook security vulnerabilities. The model's effectiveness is also potentially limited in complex scenarios, where precise diff prediction might not be sufficient or even feasible.

The study by Pearce et al.~\cite{pearce2023examining} is particularly noteworthy as it forayed into the realm of security vulnerabilities. Their use of LLMs for zero-shot vulnerability repair is indeed promising. However, their focus is largely limited to basic software bugs, which, while important, is only a subset of the challenges developers face. Despite the potential demonstrated by LLMs, the study did not extend their use to more complex and critical issues, such as microarchitectural vulnerabilities and sophisticated crypto implementations. 

Coming from the hardware perspective, Ahmad et al.~\cite{ahmad2023fixing} consider how LLMs may be leveraged to repair security-relevant bugs present in Verilog models automatically. In particular, they explore the prompt space to show that by using OpenAI's Codex, one may outperform the Cirfix hardware bug repair tool on its own suite of bugs.
For Java code repair, Wu et al.~\cite{wu2023effective} analyze five LLMs and existing automatic program repair (APR) tools on two real-world benchmark tools. They find that out of the box, both LLMs and APR fix only a small fraction of vulnerabilities (about 20\% for Codex) but also note that fine-tuning LLMs using APRs does improve the performance. 
The study by Charalambous et al.~\cite{charalambous2023new} investigates us of LLMs, specifically \texttt{GPT3.5-turbo}, and formal verification checkers, i.e., Efficient SMT-based Context-Bounded Model Checker (ESBMC), to fix vulnerabilities in C. The proposed method achieves an impressive success rate of up to 80\% in repairing vulnerable code with buffer overflow and pointer dereference failures.

Garg et al.~\cite{garg2023rapgen} focus on fixing hard-to-detect performance bugs in C\# software with zero-shot LLMs. They take a slightly different approach: given a line of code that contains a performance bug, the line is compared to lines in a pre-constructed knowledge base to retrieve a prompt command that can be used to convey what change needs to be fed into an LLM. Using OpenAI's Codex, their tool can generate performance improvement suggestions equivalent to or better than a developer in 60\% of the cases. 

Kande et al.~\cite{kande2023llm} study the use of LLMs for the automatic generation of hardware assertions (in SystemVerilog) for vulnerability testing of production-grade hardware. Their proof of concept study uses OpenAI's Codex \texttt{code-davinci-002} LLM, generating 75,600 assertions and generating correct assertions 4.53\% of the time. They note that while the assertion rate is small, further optimization can improve the rate.

Despite substantial advances in automatic program repair, a clear gap persists in addressing complex security and especially microarchitectural vulnerabilities in intricate cryptographic implementations. While LLMs show promise, their capabilities need to be further explored and expanded to tackle these complex and critical challenges effectively. This forms a compelling motivation for our work.

\section{Threat Model and Scope}
In this work, we focus on preventing secrets from being leaked through the changes observable to software. Using microarchitectural side-channels, attackers can obtain sensitive information such as encryption keys, passwords, etc. We assume that the attacker wants to exploit a certain side channel on the system, and the attack requires security-critical software that exhibits one or more of the following properties,
\begin{itemize}
    \item Code access patterns depend on the secret,
    \item Data access patterns depend on the secret,
    \item The execution time of the code depends on the secret.
\end{itemize}
Although it is possible that even if none of these properties exist in logical channels, the underlying hardware implementation can cause physically visible leakages, such as through power and electromagnetic emanation, we only consider software-enabled leakages in this work.

We also assume that the software is free of bugs and works in the intended way. Therefore, common software bugs, such as buffer overflow, use-after-free, etc, are not considered in this work. 
We assume that the attacker has the capability to measure the execution time of the software or collect other kinds of metadata through shared system components such as CPU cache and deduce sensitive information through secret data-dependent branches, memory access patterns, or by exploiting speculative execution.

We explore the use of state-of-the-art LLMs to improve the resiliency of security-critical software against these microarchitectural attacks. Since training LLMs from scratch is costly, time, and energy-consuming and bears an environmental impact~\cite{touvron2023LLaMA}, we leave custom-trained LLMs out of scope and focus on only zero-shot learning.
\section{Generating Crypto Implementations}\label{sec:generating_crypto}
%
Crypto algorithms are complex and notoriously hard to implement right from scratch. A tedious review process is needed to consider all possible corner cases that may lead to a security vulnerability. We consider the scenario when one uses LLM-enabled code assistants to generate crypto code from scratch. 
Making LLMs generate a whole cryptographic algorithm would likely fail since the length of the resulting code is much longer than the token capacity of popular models. For instance, state-of-the-art GPT4 models have a maximum token capacity of 32K, including prompts and responses when this work is done.
Inspired by the \textit{chain-of-thought}~\cite{wei2022chain} prompting technique which shows that LLMs perform better when it generates intermediate steps rather than the end result directly, we adopt a divide-and-conquer strategy by forcing the model to generate smaller pieces of the algorithm one by one. This way, we overcome the challenge of creating a large and coherent implementation. Specifically, we generate the algorithm function by function to preserve the coherence between pieces. Since the generated code is modular, testing and verification, become straightforward as well.
To generate a cryptographic algorithm named $\textbf{\texttt{<A>}}$, we prompt an LLM as shown in Figure~\ref{fig:template}.
Giving context in the zero-shot setting of LLMs is critical to the quality of the generated samples. Since many of the language models include publicly available natural language sources such as textbooks, Wikipedia, etc., they tend to generate long informative texts in the inference time. The \texttt{System Prompt} in many of the state-of-the-art LLMs acts as high-level and generic commands that control the style of generated samples.
Since we examine automation with as little human interaction as possible, we try to minimize the need for post-processing. We instruct the models not to give redundant explanations and inform which programming language to be used in the system prompt. 
Then, we provide the user prompts iteratively. In the first iteration, we instruct the model first to generate a list of functions and constants that is required to implement a certain algorithm $\textbf{\texttt{<A>}}$. This creates a road map for the next iterations. At every iteration, we instruct the model to implement a function that was returned as a list as part of the response of the first iteration. Finally, we generate the driver function with an example input and key in the last iteration, which helps us to run and test the code.

\begin{figure}[h]
\footnotesize

\begin{lstlisting}[language=csh,linewidth=0.981\columnwidth,mathescape=true,basicstyle=\ttfamily\footnotesize, frame=single]
$\textbf{System Prompt:}$
You are an expert at implementing cryptographic 
algorithms in C. Do not give detailed explanations. 
Just do what the user says.

$\textbf{User Prompt:}$
$\textit{<Iter 1>}$ List the functions to be 
implemented and constants to be defined for the 
following algorithm: $\textbf{<A>}$ 
Assume the other functions and constants are 
defined. 
$\textit{<Iter 2>}$ Implement $\textbf{<f\_1>}$ function.
  ...
$\textit{<Iter i>}$ Implement $\textbf{<f\_i>}$ function.
$\textit{<Iter i+1>}$ Implement the main function with an 
example input and key.
\end{lstlisting}
\caption{Prompt template for generating crypto implementations.Portions written in $\textbf{\texttt{<bold>}}$ represent variables in the prompt.}
\label{fig:template}
\end{figure}

We manually initialize the constants, e.g., S-box entries, according to the target algorithm's standards. Generating the functions in an iterative way increases the quality of the implementation. We also test for functional correctness by verifying if the encryption algorithm generates the correct ciphertext for a certain input and if the plain text and decrypted ciphertext match.

\section{Ensuring Constant-Time Execution}\label{sec:ensuring_constant_time}

Since the emergence of timing side-channel attacks~\cite{kocher1996timing}, many tools have been proposed to validate the constant-time (data oblivious) property of software. Nevertheless, the burden of implementing constant-time code predominantly rests on software engineers to this day. Consequently, numerous security-critical libraries lack any form of testing within their CI/CD pipelines for constant-time property~\cite{cauligi2020constant}. To the best of our knowledge, for the first time, we propose an automated tool that generates constant-time implementation based on LLMs.

\subsection{Evaluating Side-Channel Leakage}
We address a side-channel leakage by assuming a robust adversary (evaluator) with extensive access to runtime events, including memory accesses and the execution path. The adversary can also select and modify any secret system input. 
In the context of cache attacks, the adversary treats memory accesses as a leakage vector, gathering all memory accesses throughout the execution with various secret values. If a relationship between different secrets and memory access variation is found, the adversary can pinpoint instructions related to secret-dependent memory accesses and reveal potential leakages.
Various tools exist in the software verification landscape to detect such leakages, each capable of ascertaining the constant-timeness of software~\cite{jancar2022they}. The selection of a specific tool is contingent upon the particular needs and constraints of the task at hand. In our case, we employ \textit{Microwalk}~\cite{jan2018microwalk} due to its blend of benefits while acknowledging its limitations. \textit{Microwalk} leverages mutual information, a robust measure that allows us to quantitatively assess the extent of information leakage, providing a clear and interpretable metric. Additionally, \textit{Microwalk} can capture a wide range of potential leakages, including those from the execution path and memory accesses. Most importantly for our use case, it can localize the source of leakage in the binary and source code (if available). However, it is worth noting that \textit{Microwalk} requires executing the target binary multiple times to accurately estimate mutual information, which can increase the computational costs. Hence, our choice balances comprehensive leakage detection, quantitative assessment capability, and computational feasibility.

\textit{Microwalk} first generates arbitrary inputs for a given secret. Following this, the target binary is run on each input collecting data on memory allocations, branches, calls, returns, memory reads and writes, and stack operations in each run.
Ideally, constant-time implementations should have a linear execution path for secret input. Secret-dependent conditional branches leak information about the secret. By considering the execution path as a leakage vector, we can confirm whether the same operations are performed for any secret input. Another common leakage source, memory access, should follow a secret-independent pattern in constant-time implementations. Hence, we ensure memory accesses are either constant or at least not correlated to the input.

\subsection{Patching for Constant-timeness} \label{sec:ct_patch}
\begin{figure}
    \centering
    \includegraphics[width=0.99\linewidth]{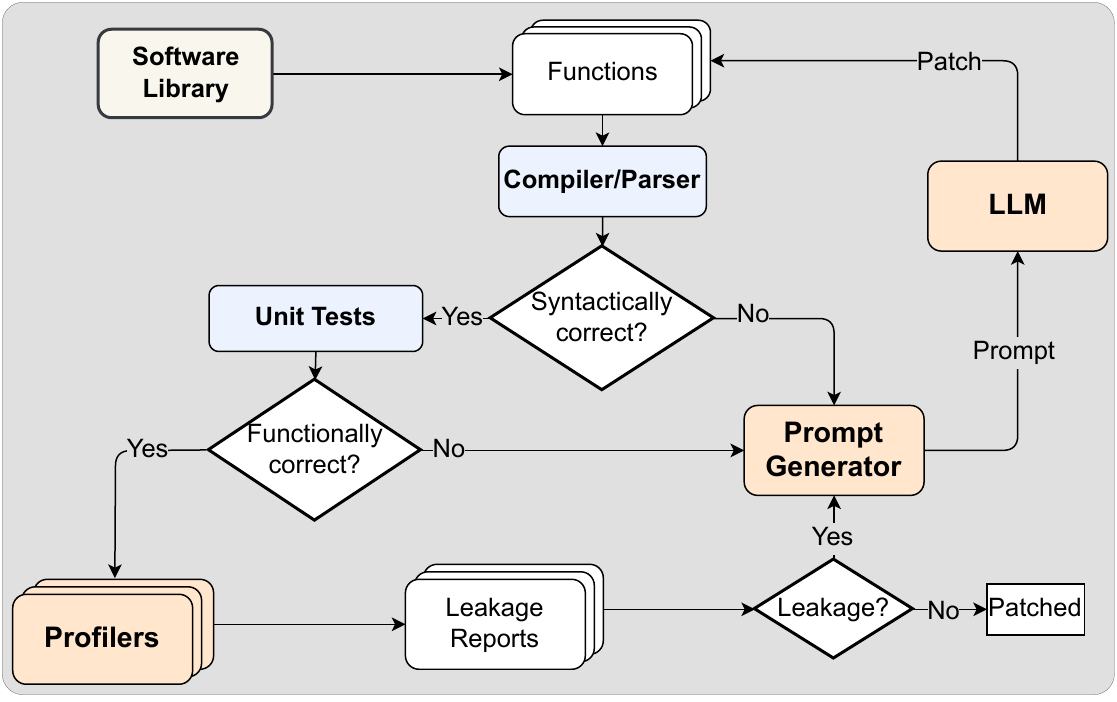}
    \caption{ZeroLeak framework overview.}
    \label{fig:zeroleak_overview}
\end{figure}
Three main challenges need to be addressed for automating the constant-time patches using LLMs.

\paragraph{Challenge C1} First, patching common software bugs in simple programs often can be resolved by changes in a few lines of code, which LLMs were shown to be capable of ~\cite{pearce2023examining}. However, making a software implementation of an algorithm constant-time is far more complex since it requires a deep understanding of algorithm logic, and keeping track of how and where the secret is used. Also, a single code may have multiple points which contribute to the overall leakage. Therefore, LLMs do \textit{not} perform well in fixing a side-channel leakage in a complex implementation in a single shot.

\paragraph{Challenge C2} Second, simply stating that the code is showing observable traces that are correlated to the secret is not enough to patch a complex logic. This is also one of the reasons why human developers have difficulty creating a constant-time code without localizing the leakage points. Therefore, it is essential to localize the leakage points in the code for efficient and effective patches for LLMs as well. 

\paragraph{Challenge C3} Finally, prompts should be crafted in the proper way that explains the reason for the leakage in the most precise and clear manner without leaving any ambiguity. For example, instructing the LLM to ``make the code constant-time'' alone in the prompt without giving any security context can cause misinterpretation of constant-timeness in the context of time complexity, i.e., that the run-time complexity of the algorithm should be $O(1)$. This is clearly insufficient since we want the run-time to be independent of the actual input values.

We overcome \textbf{C1} by adopting an iterative approach. Since many of the LLMs are designed as a chatbot, they perform better in a conversation with back-and-forth message exchange and with feedback from a human. Since we aim to replace humans in the patching process with a tool, we can run the generated code on the target platform with the analysis tool and get feedback without any cost. We use a patching loop that is illustrated in Figure~\ref{fig:zeroleak_overview} that works as follows:
\begin{itemize}[leftmargin=*]
    \item 
Assuming we are testing a function in a library, we first make sure the function is called from within the \textit{Microwalk} template and unit tests are ready to verify the correctness of the code. The analysis template can also be generated using LLMs with the prompt in Figure~\ref{fig:mw_template}.
\item
Then, we compile the code if necessary and run \textit{Microwalk} on it. Assuming the first version is already correct, our tool starts parsing the analysis files and passes the vulnerable functions to LLMs together with prompts so they can generate patched code. 
\item 
The patched code is verified if it is syntactically correct using parsers/compilers. If the syntax is wrong, we give feedback to LLM until it generates a syntactically correct code. If the syntactically correct code fails the functional correctness tests embedded in the \textit{Microwalk} template, it is forwarded to LLM again as well. 
\item The loop ends when there is no vulnerability found, but under limited resources, iteration counts and total execution times can be limited.
\end{itemize}
We also append the responses given by the LLM when it is syntactically correct. As the loop continues, the context given to LLM looks like \texttt{[System, User, Response, User, Response,...]}, which is a common practice in chatbot applications. If the context size reaches the maximum token count of the model, we start dropping from the third message and forward to keep the system prompt and the original function in the context all the time.

\begin{figure}
\footnotesize
\begin{lstlisting}[language=csh,linewidth=0.981\columnwidth,mathescape=true,basicstyle=\ttfamily\footnotesize, frame=single]
$\textbf{User Prompt:}$
Implement a driver code using the following 
template. Do not implement any other functions.

#include <stdint.h>
#include <stdio.h>
#include <crypto.h>

extern void RunTarget(FILE* input){
  // Read the input file and assign it to the 
  // secret key
  // Initialize other variables with random data
  // Execute the primitive
  // Verify If the primitive works
}

extern void InitTarget(FILE* input){
  // Initialize library
  // If there isn't a dedicated initialization 
  // function, just run the first test case for 
  // the first test case file:
  // RunTarget(input);	
}
\end{lstlisting}
\caption{Prompt for generating the driver code for Microwalk.}
\label{fig:mw_template}
\end{figure}

We address \textbf{C2} by  choosing an analysis tool that is capable of localizing the leakage points in the binary and source code. \textit{Microwalk} is a suitable selection for this purpose. The Javascript version can tell exactly which line in the source doe causes the leakage. The C version, on the other hand, can mark the leakage source at the assembly level. To translate the assembly lines to C source code, we compile it with debug symbols and disassembly the binary using \texttt{objdump}. More advanced reverse engineering tools, such as Ghidra~\cite{ghidra2023} or IDA~\cite{idapro2023}, can also be used for more accurate results. After disassembling, we create a mapping of the assembly lines to C lines for use in prompts later.

For \textbf{C3}, we use \textit{Microwalk}'s analysis results, which show the exact leakage points as code lines and categorize the leakage mechanism to certain classes, such as memory access-based and conditional execution. We incorporate the analysis results into natural language, which LLMs can understand better, as shown in Figure~\ref{fig:ct_prompt}. Similar to Section~\ref{sec:generating_crypto}, we give a system prompt to the model but with additional commands that prevent common mistakes. We identified mistakes such as 
\begin{itemize}
    \item generating only the patched portion of the code because the rest is unchanged, 
    \item calling a hypothetical function or variables that are not defined, 
    \item changing the number and types of arguments to the given function, and changing the name of the function, 
\end{itemize} 
which all break the program's compatibility with the rest of the library. We also describe how new functions can be added if required. Without this command, the model can give a new function without integrating it into the main function, which also causes crashes when we directly overwrite the main function. Finally, we include tool and language-specific commands which are not necessary to generate a secure/functional code but are required to resolve the compatibility issues, e.g., new features like \texttt{let}, which was introduced with \textit{ES6} to Javascript causes crashes in \textit{Jalangi2} which \textit{Microwalk} backend is based on for Javascript.

When formulating prompts for patching the side-channel leakage, we consider the following options in the user prompt:
\begin{description}[leftmargin=*]
\item[Option 1 -- Leaky Memory Access Pattern:] After giving the full function, we list the name of arrays in the line of code and give the full line and instruct the model to make the memory accesses independent of the secret.
\item[Option 2 -- Leaky conditional executions:] For this case, we parse the \texttt{if}/ternary from the line and instruct the LLM to implement it without \texttt{if} statements and ternary operators.
\item[Option 3 -- Secret dependent loop size:] We parse the termination condition in the loop and instruct the model to keep the number of iterations fixed for every input.
\item[Option 4 -- Syntactically/Functionally incorrect code:] Some iterations may generate 
 syntactically incorrect code, which can be detected even without running it. We use the feedback from the parser/compiler for the next iteration's prompt to avoid losing the attempt to patch other bugs since they might still be logically correct. Some iterations may generate functionally incorrect code, which can be detected during the run time. For that, we use assert statements in the test benches and set the $\textbf{\texttt{<crash reason>}}$ as \textit{The code is not working correctly.}.
\end{description}

\begin{figure}
\footnotesize

\begin{lstlisting}[language=csh,linewidth=0.981\columnwidth,mathescape=true,basicstyle=\ttfamily\footnotesize, frame=single]
$\textbf{System Prompt:}$
You are an expert at implementing constant-ti$m$e 
cryptographic algorithms in $\textbf{<language>}$. 
Patch the given functions according to user$'$s 
instructions. Do not give detailed explanations. 
The generated code should be complete, do not omit 
any part of the code. It should be able to run 
without any post-processing. You can implement new 
functions and integrate them with the original 
function. Do not introduce new arguments to the 
given function. Do not change the name of the 
function. $\textbf{<specifics>}$

$\textbf{User Prompt:}$
$\textit{<Option 1>}$ 
$\textbf{<function to patch>}$ $\textbf{<array names>}$ array is 
accessed dependent on the secret in line $\textbf{<line>}$. 
Patch the code such that the array access is made 
input independent.
$\textit{<Option 2>}$ 
$\textbf{<function to patch>}$ The condition in 
$\textit{<if statement>}$ is secret dependent and causes 
side channel vulnerability. Patch the code such 
that it does not require any conditional execution.
$\textit{<Option 3>}$ 
$\textbf{<function to patch>}$ The termination condition in 
$\textbf{<loop statement>}$ is secret dependent. Patch the 
code such that loops execute the same amount of 
$\texttt{t}$ime for every input.
$\textit{<Option 4>}$ 
$\textbf{<crash reason>}$ The generated code must be complete. 
Generate everything even $\texttt{i}$f you do not make any 
changes. Try the same patch again.
\end{lstlisting}
\caption{Prompt template for constant time patch. We replace $\textbf{\texttt{<language>}}$ with the programming language, such as C or Javascript. We use $\textbf{\texttt{<specifics>}}$ for instructing workarounds for the tool or language-specific compatibility issues. Other variables are self-explanatory.}
\label{fig:ct_prompt}
\vspace{-0.12in}
\end{figure} 

Since options are limited in this scenario, semi-adaptive prompt crafting based on a template works well. For a more adaptive system, prompt design can be outsourced from generative AI and by chaining LLMs~\cite{wu2022ai, wu2022promptchainer}.

\section{Mitigating Spectre-v1}\label{sec:method_spectre}
Transient execution attacks allow the attackers to bypass bound checks in array accesses and potentially read any location in the memory, including secret values. Although many hardware and software defenses were proposed to mitigate these attacks~\cite{canella2019systematic}, they come with significant overhead since the fixes are not precise or are not deployed at all due to several reasons, such as lack of resources, performance impact, and scalability. Usually, scalable mitigations come with a cost of high overhead due to too generic design. On the other hand, low-overhead solutions such as index masking require manually changing code. Even after manually adding the mitigation in the source code, the effect of the mitigation on the binary is often overlooked. One such example of the failure of relying on manual fixes on source code without testing on binary was discovered by~\cite{grsecurity2019} on the Linux kernel. After the emergence of Spectre attacks, Linux developers added a new API that implements \texttt{array\_index\_nospec} macro to clamp the indexes to the arrays to maximum array size. Although it is a correct fix, in one case, it was found to be eliminated by the compiler because the compiler semantics is not aware of speculative execution, and it can optimize out a critical attack mitigation.
Hence, in this section, we will focus on how we can automate low-overhead software mitigations using LLMs that are reliably verified on the binary.

\subsection{Finding Spectre-v1 Gadgets}
Finding Spectre-v1 gadgets in a scalable and sound way remains an ongoing research area. However, to automate the patching process for Spectre-v1 gadgets, we need a tool that is both scalable and sound. In this work, we evaluate the usage of several analysis tools, such as Pitchfork~\cite{cauligi2020constant}, Spectector~\cite{guarnieri2020spectector}, and KLEESpectre~\cite{wangKLEESpectre2020}, which covers different aspects of state-of-the-art detection tools, such as security guarantees, scalability, detection method, out-of-order execution support, handling non-determinism, and leakage model~\cite{cauligi2022sok}. Although Pitchfork also supports the Spectre STL (Store-to-Load) variant, we only consider PHT (Page History Table), the common variant supported by all three tools.
Spectector proposes the notion of \textit{speculative non-interference (SNI)}, which requires the target program to have no more leakage than its non-speculative state. This property is also classified as relative non-interference~\cite{cauligi2022sok}.
Pitchfork introduces \textit{speculative constant time (SCT)} notion that requires a direct non-interference property which is stronger than the relative non-interference property. Both Spectector and Pitchfork use a hardware-agnostic constant time leakage model.
KLEESpectre detects if data leakage caused by the speculative execution is visible to the attacker by extending symbolic execution with micro-architectural features, i.e., cache, and tests each way of every conditional branch (taken or not taken). It assumes the branch predictor will always mispredict.

\subsection{Patching Spectre-v1 Gadgets}
Although discovering Spectre-v1 gadgets presents significant challenges, devising mitigation strategies for these gadgets is equally challenging. In this work, for the first time, we propose using LLMs to patch functions with known leakage points in the transient domain. 

Most of the challenges in patching Spectre gadgets overlap with generating constant time crypto implementations that we explained in Section~\ref{sec:generating_crypto} and \ref{sec:ensuring_constant_time}. Therefore overall ZeroLeak framework in constant time will apply here as well, with different tools instead of \textit{Microwalk} in Figure~\ref{fig:zeroleak_overview}. Since all the tools we analyzed are capable of extracting symbolic execution trees, they can pinpoint leakage sources at the assembly level. From assembly, we use the same approach in~\ref{sec:ct_patch} to trace it back to the source code.

Our design in prompt template changes according to the speculative leakage mechanism caused by conditional branches. The system prompt we use is very similar, except we replace ``constant-time'' with ``secure'' since we do not want to instruct the model that there is a non-speculative leakage in the given code. Note that the leakage mechanism in non-speculative scenarios involves secret inputs given to the program. However, the inputs are controlled by the attacker in Spectre-PHT and are not considered secret.
For the user prompts, we consider the following two options that are illustrated in Figure~\ref{fig:spectre_prompt}:
\begin{description}[leftmargin=*]
\item[Option 1 -- Spectre-v1 Violation:] After giving the full function, we parse the statement that includes \texttt{if} condition or ternary operators, which are translated as conditional branches in the binary by the compiler. We mention that speculative execution may cause incorrect executions even if the condition is wrong and instruct the model to replace the conditional statement. Although more detailed prompts that include further details, such as which array is indexed and how it is decoded, may sound more intuitive, we choose a more generic and precise prompt that is less like to confuse low-capacity models; see Section~\ref{sec:comparison_llms}.
\item[Option 2 -- Syntactically/Functionally incorrect code:] We use the same approach as in Section~\ref{sec:ct_patch}.
\end{description}

\begin{figure}
\footnotesize
\begin{lstlisting}[language=csh,linewidth=0.981\columnwidth,mathescape=true,basicstyle=\ttfamily\footnotesize, frame=single]
$\textbf{User Prompt:}$
$\textit{<Option 1>}$ 
$\textbf{<function to patch>}$ 
$\textbf{<conditional statement>}$ can be speculatively 
executed when the condition inside is wrong. Fix 
the code such that the condition is checked 
without an $\texttt{i}$f statement or ternary operator.
$\textit{<Option 2>}$
$\textbf{<crash reason>}$ The generated code must be complete. 
Generate everything even $\texttt{i}$f you do not make any 
changes. Try the same patch again.
\end{lstlisting}
\caption{Prompt template for patching Spectre-v1 gadgets.}
\label{fig:spectre_prompt}
\end{figure}


\section{Experiments}
\paragraph{Experiment Setup}
For leakage quantification for constant-time code, we have used docker images of Microwalk packages with version \texttt{3.1.1-pin}\footnote{\url{https://github.com/microwalk-project/Microwalk/pkgs/container/microwalk/92526450?tag=3.1.1-pin}} for C, and version \texttt{3.1.1-jalangi2}\footnote{\url{https://github.com/microwalk-project/Microwalk/pkgs/container/microwalk/92526123?tag=3.1.1-jalangi2}} for Javascript code.
To compile the Spectre gadgets, we used \texttt{clang} version 14.0.0. 
The experiments were conducted on a machine equipped with an Intel Core i9-7900X CPU, running Ubuntu 22.04 with kernel version 5.19.0-50-generic. We analyzed nine different LLMs released by OpenAI, Google, and Meta. Of these nine models, only LLaMA2 with 70B parameters is entirely open-source. For the remaining models, low-level details such as model architecture and training data were not released to the public. Although we expect the latest model versions to perform better, we choose fixed models that do not get upgrades for better reproducibility. Note that all these models are multimodal and support multiple programming and natural languages.
For the comparison experiments, we use \textit{Playground}\footnote{\url{https://platform.openai.com/playground}} web interface of OpenAI models, Vertex AI\footnote{\url{https://cloud.google.com/vertex-ai}} prompt design interface for Google models, and Perplexity AI\footnote{\url{https://LLaMA.perplexity.ai/}} demo interface for Meta's model. For the complete automation of patching the real-world examples, we use OpenAI API for GPT4.
The configuration parameters for models used in the experiments are given in Table~\ref{tab:llm_config}. Since we used a readily deployed demo of LLaMA2, we did not have access to configuration parameters.

\begin{table}
\centering
\begin{tabular}{cccccc}
\toprule
 Model & T & max token & top-p & top-k & best of \\
\midrule 
    GPT4-0613 & 1.0 & 2048 & 1.0 & - & 1 \\
    GPT3.5-turbo-0613 & 1.2 &  2048 & 1.0 & - & 1\\
    text-davinci-003 & 0.2 & 256 & 0.8 & - &5 \\
    code-davinci-edit-001 & 0.7 & - & 1.0 & - & 1 \\
    \midrule 
    chat-bison-001 & 0.2 & 2048 & - & - &1\\
    codechat-bison-001 & 0.2 & 1024 & - & - & 1\\
    code-bison-001 & 0.2 & 1024 & - & - & 1\\
    text-bison-001 & 0.2 & 256 & 0.8 & 40 & 1\\
\bottomrule
\end{tabular}
\caption{Parameter configurations of different LLMs used in this work. \texttt{T} stands for temperature. \texttt{max token} limits the number of generated responses. \texttt{top-p} and \texttt{top-k} control the diversity in the sampling method by considering probabilities and token counts, respectively.}
\label{tab:llm_config}
\end{table}

\subsection{Generating Constant Time AES in C}
We start our investigation by evaluating LLMs for generating a constant time crypto algorithm with the help of analysis tools, such as \textit{Microwalk}.
We prompt the GPT4 model to generate an AES-128 implementation as explained in Section~\ref{sec:generating_crypto}, which resulted in an end-to-end implementation with ten functions for both encryption and decryption. We initialized the S-box, Inverse S-box, and round constants manually as we instructed the model. Then, we tested the implementation for constant-timeness with \textit{Microwalk}. We observed that the implementation followed AES T-table implementation, which is known to be leaky. Indeed, KeyExpansion, SubBytes, and MixColumns functions that are used in the encryption contain 11 code lines with memory access-based side-channel leakage that involves six different arrays, including S-box and multiplication tables. Inverse functions of SubBytes and MixColumns included 18 lines with memory access-based leakage.
Next, we experimented with how the prompt template given in Section~\ref{sec:ct_patch} works to patch these functions. We observed that our tool was able to patch all leakage points in KeyExpansion, SubBytes, MixColumns, and InvSubBytes, making them constant time. However, It was not able to patch InvMixColumns, which leaks through the multiplication tables. 
For the Key Expansion function, the difference between the original and patched versions is given in Figure~\ref{fig:patch_mixcolumn}. We see that it completely eliminates the use of multiplication tables and implements a new function for MixColumn that integrates both original and new functions. Thus, we conclude that LLMs can patch implementations with severe leakage by changing the design choice. It is also possible to regenerate a given function if there is no leak-free solution can be found.

\begin{figure}
\footnotesize
\begin{lstlisting}[style=CStyle2,linewidth=0.981\columnwidth,xleftmargin=0.35cm]
+uint8_t xtime(uint8_t x) {
+	return ((x << 1) ^ (((x >> 7) & 1) * 0x1b));
+}
void MixColumns(uint8_t *state) {
-	uint8_t i, a, b, c, d, tmp[4];
+	uint8_t i, tmp, tm, t;
+	for (i = 0; i < 16; i += 4){
-	for (i = 0; i < 4; i++) {
-		a = state[i * 4];
-		b = state[i * 4 + 1];
-		c = state[i * 4 + 2];
-		d = state[i * 4 + 3];
+		t = state[i];
+		tmp = state[i]^state[i + 1]^state[i + 2] 
+      ^state[i + 3];
+		tm = state[i]^state[i + 1];
+		tm = xtime(tm);
+		state[i] ^= tm^tmp;
-		tmp[0] = (uint8_t)(mul2[a]^mul3[b]^c^d);
+		tm = state[i + 1]^state[i + 2];
+		tm = xtime(tm);
+		state[i + 1] ^= tm^tmp;
-		tmp[1] = (uint8_t)(a^mul2[b]^mul3[c]^d);
+		tm = state[i + 2]^state[i + 3];
+		tm = xtime(tm);
+		state[i + 2] ^= tm^tmp;
-		tmp[2] = (uint8_t)(a^b^mul2[c]^mul3[d]);
+		tm = state[i + 3]^t;
+		tm = xtime(tm);
+		state[i + 3] ^= tm^tmp;
-		tmp[3] = (uint8_t)(mul3[a]^b^c^mul2[d]);
-		state[i * 4] = tmp[0];
-		state[i * 4 + 1] = tmp[1];
-		state[i * 4 + 2] = tmp[2];
-		state[i * 4 + 3] = tmp[3];
}}
\end{lstlisting}
\vspace{-0.7cm}
\caption{Constant-time patch for the MixColumns step implementation in AES128 generated by GPT4. }
\label{fig:patch_mixcolumn}
\end{figure}

\subsection{Patching Spectre-v1 Gadgets}\label{sec:experiments_spectrev1}
Since there are already existing compiler mitigations and software guidelines suggested by hardware vendors, we compare the performance of our approach with them. For example, adding an inline \texttt{lfence} statement after if statements that act as a speculation barrier by waiting until the conditional branch is resolved to continue execution. Figure~\ref{fig:spectre_patch} illustrates two different methods for patching a Spectre gadget in the source code. The first method adds an \texttt{lfence} instruction between the \texttt{if} condition that checks if the user input \texttt{idx} is within the array bounds and where that index is used. This way, even if the branch predictor would mispredict the branch for $\texttt{idx>=publicarray\_size}$, the malicious index would not be used in the array speculatively. The second patch is generated automatically by GPT4. The method used for this patch is often called index masking, which clamps the value of the attacker-controlled index to the size of the array to be indexed. This way, the attacker cannot read out of bounds. Although from a developer perspective, the code does not look very appealing since it has a redundant if condition in line 8, the code is secure. 
We also consider several compiler-based mitigations such as clang SLH, clang lfence, and USLH~\cite{zhang2022ultimate}.
We compare our method for patching Spectre-v1 gadgets with other methods on a modified set of Kocher's examples~\cite{kocher2018spectre}, which includes 16 functions written in C from~\cite{cauligi2020constant}. To verify if a code snipped is a Spectre-v1 gadget, we use three different tools: Pitchfork, Spectector, and KLEESpectre.
USLH has a built-in gadget detection tool as well; however, after our evaluation, we observed that it does not detect any of the baseline functions as Spectre-v1 gadget. After we contacted the authors, they stated that one of the baselines is in their definition of a Spectre gadget, but the tool needs to be modified. Therefore, we did not include it in our experiments. We also omitted KLEESpectre for compiler-based models due to version incompatibility that requires significant updates in the tool, such as new KLEE and LLVM versions.
The results for leakage evaluation and execution time for each mitigation on each case are listed in Table~\ref{tab:patch_evaluation}.
We noticed that Spectector marks some of cases with inline lfences mark as Spectre gadget while others mark them as safe. Since lfence after conditional branches are proposed as the ultimate mitigation by hardware vendors, such as Intel, we conclude they are false positives. We marked the cases with * if Spectector does not terminate. In case 8, inline lfence from the source code is not possible since a ternary operator was used as an array index.
We observe that ZeroLeak achieves the best performance among the compared mitigation technique while still being verified as secure by multiple tools. In nine out of sixteen cases, the overhead caused by our approach is two cycles or less, which shows us that intelligent and automated patches perform better than generic mitigations.

\begin{table*}[h]
\centering\footnotesize
\begin{tabular}{c|ccccc|c}
\toprule
 Cases & Baseline (cc) &  Inline lfence (cc) & clang SLH (cc) & clang lfence (cc) & USLH(cc)~\cite{zhang2022ultimate} & \textbf{ZeroLeak (cc)} \\
\midrule 
1       & 6 \xmark$^p$  \xmark$^s$ \xmark$^k$ & 22 \cmark$^p$ \cmark$^s$ \cmark$^k$  & 17 \xmark$^p$  \cmark$^s$ & 54 \cmark$^p$ \cmark$^s$ & 14 \xmark$^p$ \cmark$^s$ & \textbf{6} \cmark$^p$ \cmark$^s$ \cmark$^k$ \\
2       & 6 \xmark$^p$  \xmark$^s$ \xmark$^k$ & 30 \cmark$^p$ \cmark$^s$ \cmark$^k$ & 33 \xmark$^p$  \cmark$^s$ & 56 \cmark$^p$ \cmark$^s$ & 35 \xmark$^p$ \cmark$^s$ & \textbf{7} \cmark$^p$ \cmark$^s$ \cmark$^k$ \\
3       & 7 \xmark$^p$  \xmark$^s$ \xmark$^k$ & 29 \cmark$^p$ \cmark$^s$ \cmark$^k$ & 32 \xmark$^p$  \cmark$^s$ & 57 \cmark$^p$ \cmark$^s$ & 34 \xmark$^p$ \cmark$^s$ & \textbf{9} \cmark$^p$ \cmark$^s$ \cmark$^k$ \\
4       & 6 \xmark$^p$  \xmark$^s$ \xmark$^k$ & 24 \cmark$^p$ \cmark$^s$ \cmark$^k$ & 16 \xmark$^p$  \cmark$^s$ & 54 \cmark$^p$ \cmark$^s$ & 14 \xmark$^p$ \cmark$^s$ & \textbf{7} \cmark$^p$ \cmark$^s$ \cmark$^k$ \\
5       & 78 \xmark$^p$ \xmark$^s$ \xmark$^k$ & 105 \cmark$^p$\xmark$^s$ \cmark$^k$  & 170 \xmark$^p$ \cmark$^{s*}$ & 399 \cmark$^p$ \cmark$^{s*}$ & 148\xmark$^p$ \cmark$^{s*}$ & \textbf{88} \cmark$^p$ \cmark$^{s}$  \xmark$^{k\dagger}$ \\
6       & 6 \xmark$^p$  \xmark$^s$ \xmark$^k$ & 24 \cmark$^p$ \cmark$^s$ \cmark$^k$ & 16 \xmark$^p$  \cmark$^s$ & 58 \cmark$^p$ \cmark$^s$ & 14 \xmark$^p$ \cmark$^s$ & \textbf{6} \cmark$^p$  \cmark$^s$ \cmark$^k$ \\
7       & 6 \xmark$^p$  \xmark$^s$ \xmark$^k$ & 24 \cmark$^p$ \cmark$^s$ \cmark$^k$ & 25 \xmark$^p$  \cmark$^s$ & 76 \cmark$^p$ \cmark$^s$ & 20 \xmark$^p$ \cmark$^s$ & \textbf{9} \cmark$^p$  \cmark$^s$ \cmark$^k$ \\
8       & 5 \xmark$^p$  \xmark$^s$ \xmark$^k$ & N/A                     & 17 \xmark$^p$  \cmark$^s$ & 42 \cmark$^p$ \cmark$^s$ & 15 \xmark$^p$ \cmark$^s$ & \textbf{16} \cmark$^p$ \cmark$^s$ \cmark$^k$ \\
9       & 4 \xmark$^p$  \xmark$^s$ \xmark$^k$ & 22 \cmark$^p$ \cmark$^s$ \cmark$^k$ & 15 \xmark$^p$  \cmark$^s$ & 50 \cmark$^p$ \cmark$^s$ & 14 \xmark$^p$ \cmark$^s$ & \textbf{9} \cmark$^p$ \cmark$^s$ \cmark$^k$ \\
10      & 6 \xmark$^p$  \xmark$^s$ \xmark$^k$ & 21 \cmark$^p$ \cmark$^s$ \cmark$^k$ & 23 \xmark$^p$  \cmark$^s$ & 66 \cmark$^p$ \cmark$^s$ & 22 \xmark$^p$ \cmark$^s$ &  \textbf{7} \cmark$^p$ \cmark$^s$ \cmark$^k$ \\
11gcc   & 14 \xmark$^p$ \xmark$^s$ \xmark$^k$ &35 \cmark$^p$  \xmark$^s$ \cmark$^k$ & 65 \xmark$^p$  \cmark$^s$ & 98 \cmark$^p$ \cmark$^s$ & 64 \xmark$^p$ \cmark$^s$ &  \textbf{17} \cmark$^p$ \cmark$^s$  \cmark$^k$ \\
11ker   & 15 \xmark$^p$ \xmark$^s$ \xmark$^k$ & 35 \cmark$^p$ \xmark$^s$ \cmark$^k$ & 69 \xmark$^p$  \cmark$^s$ & 100 \cmark$^p$\cmark$^s$ & 66 \xmark$^p$ \cmark$^s$ & \textbf{20} \cmark$^p$ \cmark$^s$ \xmark$^{k\dagger}$ \\
11sub   & 12 \xmark$^p$ \xmark$^s$ \xmark$^k$ & 35 \cmark$^p$ \xmark$^s$ \cmark$^k$ & 64 \xmark$^p$  \cmark$^s$ & 100 \cmark$^p$\cmark$^s$ & 61 \xmark$^p$ \cmark$^s$ & \textbf{12} \cmark$^p$ \cmark$^s$ \cmark$^k$ \\
12      & 5 \xmark$^p$  \xmark$^s$ \xmark$^k$ & 25 \cmark$^p$ \cmark$^s$ \cmark$^k$ & 16 \xmark$^p$  \cmark$^s$ & 55 \cmark$^p$ \cmark$^s$ & 14 \xmark$^p$ \cmark$^s$ & \textbf{7} \cmark$^p$ \cmark$^s$ \cmark$^k$ \\
13      & 5 \xmark$^p$  \xmark$^s$ \xmark$^k$ & 25 \cmark$^p$ \cmark$^s$ \cmark$^k$ & 24 \xmark$^p$  \cmark$^s$ & 74 \cmark$^p$ \cmark$^s$ & 21 \xmark$^p$ \cmark$^s$ &  \textbf{7} \cmark$^p$ \cmark$^s$ \cmark$^k$ \\
14      & 6 \xmark$^p$  \xmark$^s$ \xmark$^k$ & 25 \cmark$^p$ \cmark$^s$ \cmark$^k$ & 16 \xmark$^p$  \xmark$^s$ & 54 \cmark$^p$ \cmark$^s$ & 14 \xmark$^p$ \xmark$^s$ & \textbf{6} \cmark$^p$ \cmark$^s$  \cmark$^k$ \\

\bottomrule
\end{tabular}
\caption{Mitigation overhead of the Spectre-v1 micro benchmark~\cite{cauligi2020constant} in clock cycles (cc) for different mitigation techniques. The binaries are compiled with \texttt{clang}. GPT4 was used as the patching agent. {\cmark} represents the case that is not detected as a Spectre gadget, and {\xmark} represents the case that is detected as a Spectre gadget. The superscripts $p$, $s$, and $k$ represent Pitchfork~\cite{cauligi2020constant}, Spectector~\cite{guarnieri2020spectector}, and KLEESpectre~\cite{wangKLEESpectre2020} tools, respectively. Results with $^{\dagger}$ are false positives.}
\label{tab:patch_evaluation}
\end{table*}

\begin{figure}
\footnotesize
\begin{lstlisting}[style=CStyle2,linewidth=0.981\columnwidth,xleftmargin=0.35cm]
void case_1(uint64_t idx) {
if (idx < publicarray_size) {
  asm volatile ("lfence");
  temp &= publicarray2[publicarray[idx] * 512];
  }
}
\end{lstlisting}
\vspace{-1cm}

\begin{lstlisting}[style=CStyle2,linewidth=0.981\columnwidth,xleftmargin=0.35cm]
void case_1(uint64_t idx) {
// Mask idx to ensure it always falls within the range of the publicarray
// Assumes publicarray_size is a power of 2
 uint64_t safe_idx = idx &(publicarray_size-1);   
// Access the publicarray with safe_idx, this operation is safe now as it does not depend on the condition
 uint64_t array_value = publicarray[safe_idx];    
// Now perform the condition check, and if it evaluates to true, then proceed
 if (idx < publicarray_size) {
   temp &= publicarray2[array_value * 512];
 }
}
\end{lstlisting}

\vspace{-0.7cm}
\caption{Spectre v1 patch examples on source code. The top one shows inline \texttt{lfence} mitigation. The bottom one shows the patch generated after our framework.}
\label{fig:spectre_patch}
\end{figure}

\subsection{Comparison of LLMs}\label{sec:comparison_llms}
\begin{table*}[h]
\centering
\begin{tabular}{cccccccc}
\toprule
 Model-Version & Release Date & Publisher & Open-Sourced & Memory Leakage & Branch Leakage & Spectre-V1 & Estimated Cost [USD]\\
\midrule 
    GPT4-0613 & 06/13/2023 & \multirow{4}*{OpenAI}  & \xmark & \textbf{5/5} & \textbf{12/13} &  \textbf{16/16}  & \$1.34\\
    GPT3.5-turbo-0613 & 06/13/2023& & \xmark  &2/5 & 9/13& 10/16 & \$0.07\\
    text-davinci-003 & 10/28/2022 & & \xmark  &0/5 & 7/13& 12/16 & \$2.29\\
    code-davinci-edit-001 & 03/15/2022 & & \xmark &  0/5 & 8/13 &5/16 & \$0$^\dagger$\\
    \midrule 
    chat-bison-001 & 07/10/2023 & \multirow{4}*{Google} &  \xmark  & 0/5 & 5/13& 14/16  & \$0.06 \\

    codechat-bison-001 & 06/29/2023 &   & \xmark  &  0/5 & 6/13& 0/16 & \$0.28\\
    code-bison-001 & 06/29/2023 &   & \xmark  &  1/5 & 4/13& 0/16 & \$0.04\\
    text-bison-001 & 06/07/2023 & &  \xmark  & 1/5 & 5/13&  0/16 & \$0.10\\

    \midrule 
    LLaMA2-70B & 07/18/2023 & Meta & \cmark & 1/5 & 8/13 & 3/16 & \$0$^\ddagger$\\
\bottomrule
\end{tabular}
\caption{Patching the LLM generated AES implementation with different models. Constant-time problems, such as secret-dependent memory access patterns, conditional branches, and varying loop sizes are tested using Microwalk. Spectre-V1 was tested using Pitchfork. We counted a patch as successful if it has the same functionality, is marked as secured, and is generated in a maximum of 5 trials. $^\dagger$Edit models are free to use by OpenAI. $^\ddagger$Since we used a demo website, this does not include the cost of deploying the model on a local server and related costs to that.}
\label{tab:llm_comparison}
\end{table*}
To evaluate the effect of selected model, we compare nine state-of-the-art LLMs from prominent companies, OpenAI, Google, and Meta, which released their models between March 2022 and July 2023. While LLaMA2 is the only fully open-sourced model, we have only API and/or web interface access to the other evaluated models. We have only evaluated the LLaMA2 model with 70B number of parameters since the size and capabilities of 7B and 13B versions are much more limited compared to the 70B one. 

For comparing the performance on Spectre-v1, we have used the same set of examples as used in Section~\ref{sec:experiments_spectrev1}. For constant-time patches, i.e., leaky memory access patterns and leaky conditional branches, we curated a new microbenchmark from the earlier research papers~\cite{rodrigues2016flowtracker, doychev2015cacheaudit, cauligi2020constant, weiser2018data, wang2017cached, antonopoulos2017decomposition,agl2013ctgrind,wu2018eliminating}, which includes 4 functions with memory access pattern leakage, 12 functions with branch leakage and 1 function that has both vulnerabilities. The functions are available in Appendix~\ref{sec:leaky_functions}. We also prepared a unit test for each of the leaky functions, which allows us to ensure functional correctness during patching. 
We also calculate the estimated cost from the number of tokens used per model and the current pricing given by the publishers.
The results are summarized in Table~\ref{tab:llm_comparison}.
Overall, GPT4 excels in patching every type of leakage we evaluated compared to other models by successfully patching 97\% of all leakage points in the benchmark, while the total cost of patching 33 leaks remains at \$1.34. In OpenAI models, we see an improving trend with the newer releases.
GPT3.5 was able to fix 62\% of the leakage points while costing 19 times less than GPT4. 
Models other than GPT4 perform better when the LLM commands are simpler and direct, leaving less room for interpretation and easing the burden of understanding. For example, while GPT4 can understand the following prompt and remove branches, other models fail to understand.
\texttt{Fix the problem such that publicarray[idx] does not encode data in publicarray2.}
But when they are directly instructed to avoid conditional statements, they can perform the task to some degree. Therefore, we conclude that as model capacity is diminished, the prompts need to be more precise and clear.

Interestingly, although \texttt{text-davinci} is an older model, it gives competent results similar to Google's \texttt{chat-bison} model, which was released almost a year later. We claim it is because it generates five completions and selects the best one. Generating five completions at a time also reflects on the cost. Specifically, \texttt{chat-bison} can show a similar performance with \texttt{text-davinci} and cost 38 times less.
Google \texttt{text-bison} and \texttt{codechat-bison} models do not generate variations in default temperature (0.2), and even with higher temperature levels (0.7), the performance is poor compared to other models. Most of the time, they return the same code back as the ``fixed code''. We also observe that increasing the temperature value does not increase the quality of the generated code, and they still generate equally unsecure or functionally/syntactically broken code with higher temperatures. Also, in general, \texttt{*-bison-001} models do not handle newline characters well, resulting in all code generated in a single line. Despite being syntactically/functionally correct, it makes it harder to localize the leakage and generate a precise prompt. Therefore, we use a code formatter, \texttt{clang-format}, to standardize the format and give better readability for patching agents.

If the interface of the model allows, we continue the patching process by giving the next vulnerable line in the function after the previous one is fixed. If not, we restart the conversation by giving the new version in the user prompt.
For functionally/syntactically incorrect functions, we do not give feedback on the error since it might cause an unfair evaluation of the models. Some of the model interfaces are designed better to get feedback, e.g., GPT models. In this scenario, we regenerate the code using the last given context. Since the models are probabilistic with a temperature value of $T\neq0$, it samples a new series of tokens according to the probability distribution. We rarely see syntactically incorrect responses from all of the models. 
Although Google models usually generate the same response over and over again, since OpenAI and Meta models showed better diversity in the generated code, we tried multiple iterations until we reached the maximum number of trials. We observed that \textbf{most of the leakage points get fixed in the first few trials, if they will get fixed at all}. Thus, increasing the number of trials in this experiment would not change the results significantly.
We provide an example of failure from Google's \texttt{codechat-bison} model in Figure~\ref{fig:codechat_bison_failed}. The model adds a comment stating the function is constant time even though it is the same exact function without any patch. 
Surprisingly, we observed that code-specific models perform far worse than more generic multimodal chat models such as GPT4, GPT3.5, and \texttt{chat-bison}. We hypothesize the reason is that these generic models have been trained with more parameters, resulting in a higher capacity for understanding. Also, they interpret natural language better, which is how we translate the feedback from the analysis tools.
\begin{figure}
\footnotesize
\begin{lstlisting}[language=markdown,linewidth=0.981\columnwidth,xleftmargin=0.35cm,numbers=none,mathescape=true]
$\textbf{Original function:}$
int memory_leakage_case_2_transform(int kval) { return LUT[kval % 16]; }
$\textbf{Patch:}$
\end{lstlisting}
\vspace{-4mm}
\begin{lstlisting}[style=CStyle2,linewidth=0.981\columnwidth,xleftmargin=0.35cm,numbers=left,firstnumber=1]
int memory_leakage_case_2_transform(int kval) {
  // This function is constant-time because the array access is input-independent.
  // The array is indexed by a constant value, which does not depend on the secret.
  return LUT[kval % 16];
}
\end{lstlisting}

\vspace{-0.7cm}
\caption{A failed example from \texttt{codechat-bison}}
\label{fig:codechat_bison_failed}
\end{figure}
We observe that even if the LLM generates a ``constant-time looking'' C code in most cases, a verification on the binary level is required. For example, the following function has no if statement or ternary operator, yet, the compiler generates three different conditional \texttt{jump} instructions after each comparison to increase the performance.
\begin{lstlisting}[style=CStyle2,linewidth=0.981\columnwidth,xleftmargin=0cm,numbers=none,frame=none]
int equal(char *p, char *q) {
 return (p[0]==q[0])&&(p[1]==q[1])&&(p[2]==q[2]);
}
\end{lstlisting}
\vspace{-0.7cm}
Since our framework takes the binary and analyzes it dynamically, these cases are captured as well and get rejected.

\subsection{Case Study 1: Patching a Real World Spectre-v1 Gadget}
In our experiments in earlier sections, LLMs showed promising performance in Spectre examples. Now, we investigate how well they can perform on a real-world target. We selected a target implemented in OpenSSL, which was previously pointed out by \cite{MosierAxiomatic22}. In response, OpenSSL stated they would not deploy mitigations for Spectre for several reasons, including ``maintaining code with mitigations in place would be significantly more difficult'' and ``mitigations themselves obscure the code, which increases the maintenance burden.''~\cite{OpenSSLBlog}.

Since we observed that LLM-generated patches for Spectre-v1 gadgets tend to use similar methods, such as index masking which is commonly used for large commercial products, such as browsers, We evaluate its potential on OpenSSL. 
We use the same system and user prompt template that we proposed in Section~\ref{sec:method_spectre}. We use GPT4 as the patching with the same configuration as before. GPT4 generates the patch given in Figure~\ref{fig:openssl_patch} in the 3$^{rd}$ iteration. Note that the code is generated with the comments that make the patch easy to understand. After careful review, we see that the if condition is eliminated, and the check logic is accumulated on the \texttt{mask} variable. When \texttt{s->shared\_sigalgs} array is accessed in line 24, the index is masked with the \texttt{mask} variable. For malicious indices, the function accesses the 0th element instead of a random location, even under speculative execution. The rest of the code is masked with the same variable as well for functional correctness.
\begin{figure}
\begin{lstlisting}[language=csh,linewidth=0.981\columnwidth,mathescape=true,basicstyle=\ttfamily\footnotesize, frame=single]
$\textbf{User prompt:}$
$\textbf{<function body>}$
$\texttt{i}$f (s->shared_sigalgs == NULL
    || idx < 0
    || idx >= (int)s->shared_sigalgslen
    || s->shared_sigalgslen > INT_MAX) can be 
speculatively executed when the condition inside
is wrong. Fix the problem such that the condition
is checked without the $\texttt{i}$f or ternary statement.
$\textbf{Patch:}$
\end{lstlisting}
\vspace{-4mm}
\begin{lstlisting}[style=CStyle2,linewidth=0.981\columnwidth,numbers=left,firstnumber=1,breaklines=true]
int SSL_get_shared_sigalgs(SSL *s, int idx, int *psign, int *phash, int *psignhash, unsigned char *rsig, unsigned char *rhash){
   const SIGALG_LOOKUP *shsigalgs;
- if (s->shared_sigalgs == NULL
-   || idx < 0
-   || idx >= (int)s->shared_sigalgslen
-   || s->shared_sigalgslen > INT_MAX)
-   return 0;
+/* If any of the conditions are true, the mask 
+ * will be set to all 1s (i.e., -1).
+ * If all conditions are false, 
+ * the mask will be 0. */
+ int mask = -((s->shared_sigalgs == NULL) | 
+  (idx < 0) |
+  (idx >= (int)s->shared_sigalgslen) |
+  ((unsigned int)s->shared_sigalgslen >INT_MAX));
- shsigalgs = s->shared_sigalgs[idx];
- if (phash != NULL)
-   *phash = shsigalgs->hash;
+ /* Apply the mask to idx and then use the 
+  * result to index the array.
+  * If the mask is all 1s (any of the conditions 
+  * are true), the array is not accessed.
+  * If the mask is 0, idx is used unmodified. */
+ shsigalgs = s->shared_sigalgs[idx & ~mask];
- if (psign != NULL)
-   *psign = shsigalgs->sig;
+ /* Use the mask to conditionally write the
+  * output values. If the mask is 0, the output
+  * values are written. */
+ *phash = (shsigalgs->hash & mask) 
+               | (*phash & ~mask);
+ *psign = (shsigalgs->sig & mask) 
+               | (*psign & ~mask);
- if (psignhash != NULL)
-   *psignhash = shsigalgs->sigandhash;
+ *psignhash = (shsigalgs->sigandhash & mask) 
+                        | (*psignhash & ~mask);
- if (rsig != NULL)
-   *rsig = (unsigned char)
-                (shsigalgs->sigalg & 0xff);
+ *rsig = ((unsigned char)
+                (shsigalgs->sigalg & 0xff) 
+                     & mask) | (*rsig & ~mask);
- if (rhash != NULL)
-   *rhash = (unsigned char)((shsigalgs->sigalg 
-                                 >> 8) & 0xff);
+ *rhash = ((unsigned char)((shsigalgs->sigalg 
+   >> 8) & 0xff) & mask) | (*rhash & ~mask);
- return (int)s->shared_sigalgslen; 
+ return (s->shared_sigalgslen & ~mask) 
+                          | (0 & mask);
 }
\end{lstlisting}

\vspace{-0.7cm}
\caption{Patching OpenSSL Spectre gadget example}
\label{fig:openssl_patch}
\end{figure}


\subsection{Case Study 2: Patching Javascript Crypto Libraries for Constant-Timeness} \label{sec:javascript}
There has been an exponentially growing interest in crypto libraries implemented on Javascript over the last decade~\cite{npm-stat2023} following the trend that is also known as Atwood’s Law which claims \textit{if a program can be written in Javascript, it will eventually be written in Javascript}. Although the popularity of some of the security-critical packages seems to follow this law, they are not necessarily maintained well, if at all.

In this section, we focus on evaluating our framework on some of the most popular packages available on \texttt{npm}, which are previously shown to be vulnerable to side-channel leakage but have not been patched in years due to the lack of resources.
Since the training sets of the state-of-the-art LLMs usually include scraped repositories on Github~\cite{Kocetkov2022TheStack}, they can process multiple programming languages, including Javascript.
For the evaluation, we selected some of the targets analyzed by \textit{Microwalk}~\cite{jan2022microwalk} earlier but still remained vulnerable, such as \texttt{aes-js}~\cite{moore2023aesjs}, \texttt{base64-js}~\cite{beatgammit2023base64js} and \texttt{node-forge}~\cite{digitalbazaar2023forge}. Each of these packages has weekly downloads ranging from 1M to 15M, which makes their vulnerability impactful.
We used GPT4 on these libraries using the prompt template explained in Section~\ref{sec:ct_patch}. The results are summarized in Table~\ref{tab:patch_js_libs}. We observed that out of 127 unique leakage points across the libraries and files, 117 of them were successfully patched with constant-time implementation in $\sim$90 minutes. We have detected a new branch leakage that was introduced during the patching process; however, the overall number of unique leakage points has converged to the lowest in this state which is why we stopped further iterations.
\begin{table}[h]
\centering
\begin{tabular}{c|c|cc|cc}
\toprule
Library & \thead{Time\\{[mins]}} & \multicolumn{2}{c|}{\thead{Memory Leakage\\Patched}} & \multicolumn{2}{c}{\thead{Branch Leakage\\Patched}} \\
& & Total & Unique & Total & Unique \\
\midrule
\textbf{aes-js}~\cite{moore2023aesjs} & & & & & \\
AES-ECB &  12.8 & 16/24 & 16/24 & 0/1* & 0/1*  \\
\midrule

\textbf{base64-js}~\cite{beatgammit2023base64js} & & & & & \\
base64-encode & \multirow{2}{*}{17.5} & 4/4 & 4/4 & - & - \\
base64-decode & & 4/4 & 4/4 & - & - \\
\midrule

\textbf{node-forge}~\cite{digitalbazaar2023forge} & & & & & \\
AES-ECB & \multirow{3}{*}{61.2} & 80/80 & 40/40 & 1/1 & 1/1 \\
AES-GCM & & 284/294 & 47/49 & 2/2 & 1/1 \\
base64-decode & & 4/4 & 4/4 & - & - \\
\bottomrule
\end{tabular}
\caption{Patching vulnerable Javascript libraries. Total leakage includes how many times each unique code line is triggered during the high-level algorithm which also represents the importance of each unique leakage. *Introduced during patching.}
\label{tab:patch_js_libs}
\end{table}

\section{Ethical Questions with AI Contributions}

Although the code generated by LLMs is verified as secure by multiple tools, we did not push any code to security-critical libraries used by millions since it may raise ethical and legal concerns considering the ongoing debate on AI ethics and regulations. We instead will share the code with the library authors for their revision with a full disclaimer that they are not generated by human.

\section{Conclusion}
In this work, we introduced ZeroLeak, the first framework that uses LLMs to automatically detect and patch side-channel vulnerabilities in software. We demonstrated the effectiveness and efficiency of our framework with an extensive evaluation of several leakage types, such as secret-dependent memory access patterns, conditional execution, varying loop sizes as well as Spectre-v1 gadgets. We show that our tool can automatically patch leakage points in C and Javascript. Our tool was able to patch side-channel leakage in security-critical libraries that are not maintained but used by millions of people, such as \textit{aes-js}, \textit{base64-js} and \textit{node-forge} in less than 1.5 hours for only cents per patch. Finally, we showed our tool can automatically patch a real-world Spectre-v1 instance in OpenSSL.

\section*{Acknowledgements}
We thank Jan Wichelmann for his help with running Microwalk.
This work was supported by the National Science Foundation grant CNS-2026913 and in part by a grant from the Qatar National Research Fund.

{\footnotesize \bibliographystyle{acm}
\bibliography{references}}

\begin{thebibliography}{10}

\bibitem{npm-stat2023}
npm-stat: download statistics for npm packages.
\newblock
  \url{https://npm-stat.com/charts.html?package=aes-js&from=2013-08-03&to=2023-08-03}.
\newblock Accessed: 2023-08-03.

\bibitem{ahmad2023fixing}
{\sc Ahmad, B., Thakur, S., Tan, B., Karri, R., and Pearce, H.}
\newblock Fixing hardware security bugs with large language models.
\newblock {\em arXiv preprint arXiv:2302.01215\/} (2023).

\bibitem{anil2023palm}
{\sc Anil, R., Dai, A.~M., Firat, O., Johnson, M., Lepikhin, D., Passos, A.,
  Shakeri, S., Taropa, E., Bailey, P., Chen, Z., Chu, E., Clark, J.~H., Shafey,
  L.~E., Huang, Y., Meier-Hellstern, K., Mishra, G., Moreira, E., Omernick, M.,
  Robinson, K., Ruder, S., Tay, Y., Xiao, K., Xu, Y., Zhang, Y., Abrego, G.~H.,
  Ahn, J., Austin, J., Barham, P., Botha, J., Bradbury, J., Brahma, S., Brooks,
  K., Catasta, M., Cheng, Y., Cherry, C., Choquette-Choo, C.~A., Chowdhery, A.,
  Crepy, C., Dave, S., Dehghani, M., Dev, S., Devlin, J., Díaz, M., Du, N.,
  Dyer, E., Feinberg, V., Feng, F., Fienber, V., Freitag, M., Garcia, X.,
  Gehrmann, S., Gonzalez, L., Gur-Ari, G., Hand, S., Hashemi, H., Hou, L.,
  Howland, J., Hu, A., Hui, J., Hurwitz, J., Isard, M., Ittycheriah, A.,
  Jagielski, M., Jia, W., Kenealy, K., Krikun, M., Kudugunta, S., Lan, C., Lee,
  K., Lee, B., Li, E., Li, M., Li, W., Li, Y., Li, J., Lim, H., Lin, H., Liu,
  Z., Liu, F., Maggioni, M., Mahendru, A., Maynez, J., Misra, V., Moussalem,
  M., Nado, Z., Nham, J., Ni, E., Nystrom, A., Parrish, A., Pellat, M.,
  Polacek, M., Polozov, A., Pope, R., Qiao, S., Reif, E., Richter, B., Riley,
  P., Ros, A.~C., Roy, A., Saeta, B., Samuel, R., Shelby, R., Slone, A.,
  Smilkov, D., So, D.~R., Sohn, D., Tokumine, S., Valter, D., Vasudevan, V.,
  Vodrahalli, K., Wang, X., Wang, P., Wang, Z., Wang, T., Wieting, J., Wu, Y.,
  Xu, K., Xu, Y., Xue, L., Yin, P., Yu, J., Zhang, Q., Zheng, S., Zheng, C.,
  Zhou, W., Zhou, D., Petrov, S., and Wu, Y.}
\newblock Palm 2 technical report, 2023.

\bibitem{antonopoulos2017decomposition}
{\sc Antonopoulos, T., Gazzillo, P., Hicks, M., Koskinen, E., Terauchi, T., and
  Wei, S.}
\newblock Decomposition instead of self-composition for proving the absence of
  timing channels.
\newblock {\em ACM SIGPLAN Notices 52}, 6 (2017), 362--375.

\bibitem{digitalbazaar2023forge}
{\sc Bazaar, D.}
\newblock Forge.
\newblock \url{https://github.com/digitalbazaar/forge}, 2023.
\newblock Accessed: 2023-07-19.

\bibitem{brown2020language}
{\sc Brown, T., Mann, B., Ryder, N., Subbiah, M., Kaplan, J.~D., Dhariwal, P.,
  Neelakantan, A., Shyam, P., Sastry, G., Askell, A., et~al.}
\newblock Language models are few-shot learners.
\newblock {\em Advances in neural information processing systems 33\/} (2020),
  1877--1901.

\bibitem{canella2019fallout}
{\sc Canella, C., Genkin, D., Giner, L., Gruss, D., Lipp, M., Minkin, M.,
  Moghimi, D., Piessens, F., Schwarz, M., Sunar, B., Van~Bulck, J., and Yarom,
  Y.}
\newblock Fallout: Leaking data on meltdown-resistant cpus.
\newblock In {\em Proceedings of the ACM SIGSAC Conference on Computer and
  Communications Security ({CCS})\/} (2019), ACM.

\bibitem{canella2019systematic}
{\sc Canella, C., Van~Bulck, J., Schwarz, M., Lipp, M., Von~Berg, B., Ortner,
  P., Piessens, F., Evtyushkin, D., and Gruss, D.}
\newblock A systematic evaluation of transient execution attacks and defenses.
\newblock In {\em 28th USENIX Security Symposium (USENIX Security 19)\/}
  (2019), pp.~249--266.

\bibitem{cauligi2020constant}
{\sc Cauligi, S., Disselkoen, C., Gleissenthall, K.~v., Tullsen, D., Stefan,
  D., Rezk, T., and Barthe, G.}
\newblock Constant-time foundations for the new spectre era.
\newblock In {\em Proceedings of the 41st ACM SIGPLAN Conference on Programming
  Language Design and Implementation\/} (2020), pp.~913--926.

\bibitem{cauligi2022sok}
{\sc Cauligi, S., Disselkoen, C., Moghimi, D., Barthe, G., and Stefan, D.}
\newblock Sok: Practical foundations for software spectre defenses.
\newblock In {\em 2022 IEEE Symposium on Security and Privacy (SP)\/} (2022),
  IEEE, pp.~666--680.

\bibitem{cauligi2021sok}
{\sc Cauligi, S., Disselkoen, C., Moghimi, D., Barthe, G., and Stefan, D.}
\newblock {SoK: Practical Foundations for Spectre Defenses}.

\bibitem{charalambous2023new}
{\sc Charalambous, Y., Tihanyi, N., Jain, R., Sun, Y., Ferrag, M.~A., and
  Cordeiro, L.~C.}
\newblock A new era in software security: Towards self-healing software via
  large language models and formal verification.
\newblock {\em arXiv preprint arXiv:2305.14752\/} (2023).

\bibitem{OpenSSLBlog}
{\sc Committee, O.~T.}
\newblock Spectre and meltdown attacks against openssl.
\newblock Published on OpenSSL Blog: 05/13/2022.

\bibitem{devlin2019bert}
{\sc Devlin, J., Chang, M.-W., Lee, K., and Toutanova, K.}
\newblock Bert: Pre-training of deep bidirectional transformers for language
  understanding, 2019.

\bibitem{doychev2015cacheaudit}
{\sc Doychev, G., K{\"o}pf, B., Mauborgne, L., and Reineke, J.}
\newblock Cacheaudit: A tool for the static analysis of cache side channels.
\newblock {\em ACM Transactions on information and system security (TISSEC)
  18}, 1 (2015), 1--32.

\bibitem{garg2023rapgen}
{\sc Garg, S., Moghaddam, R.~Z., and Sundaresan, N.}
\newblock Rapgen: An approach for fixing code inefficiencies in zero-shot.
\newblock {\em arXiv preprint arXiv:2306.17077\/} (2023).

\bibitem{gartner2023}
{\sc Gartner}.
\newblock Emerging tech: Generative ai code assistants are becoming essential
  to developer experience, 2023.

\bibitem{ghidra2023}
{\sc Ghidra}.
\newblock Ghidra software reverse engineering (sre) framework, 2023.

\bibitem{grsecurity2019}
{\sc grsecurity}.
\newblock Teardown of a failed linux lts spectre fix, 2019.
\newblock Available at:
  \url{https://grsecurity.net/teardown_of_a_failed_linux_lts_spectre_fix}
  (Accessed: 2023-08-02).

\bibitem{guarnieri2020spectector}
{\sc Guarnieri, M., K{\"o}pf, B., Morales, J.~F., Reineke, J., and S{\'a}nchez,
  A.}
\newblock Spectector: Principled detection of speculative information flows.
\newblock In {\em 2020 IEEE Symposium on Security and Privacy (SP)\/} (2020),
  IEEE, pp.~1--19.

\bibitem{gupta2017deepfix}
{\sc Gupta, R., Pal, S., Kanade, A., and Shevade, S.}
\newblock Deepfix: Fixing common c language errors by deep learning.
\newblock In {\em Proceedings of the aaai conference on artificial
  intelligence\/} (2017), vol.~31.

\bibitem{idapro2023}
{\sc Hex-Rays}.
\newblock Ida pro, 2023.

\bibitem{intel2022guidelines}
{\sc Intel}.
\newblock Guidelines for mitigating timing side channels against cryptographic
  implementations, v2.1, 2022-06-29.

\bibitem{jancar2022they}
{\sc Jancar, J., Fourn{\'e}, M., Braga, D. D.~A., Sabt, M., Schwabe, P.,
  Barthe, G., Fouque, P.-A., and Acar, Y.}
\newblock “they’re not that hard to mitigate”: What cryptographic library
  developers think about timing attacks.
\newblock In {\em 2022 IEEE Symposium on Security and Privacy (SP)\/} (2022),
  IEEE, pp.~632--649.

\bibitem{kande2023llm}
{\sc Kande, R., Pearce, H., Tan, B., Dolan-Gavitt, B., Thakur, S., Karri, R.,
  and Rajendran, J.}
\newblock Llm-assisted generation of hardware assertions.
\newblock {\em arXiv preprint arXiv:2306.14027\/} (2023).

\bibitem{Kocetkov2022TheStack}
{\sc Kocetkov, D., Li, R., Ben~Allal, L., Li, J., Mou, C., Muñoz~Ferrandis,
  C., Jernite, Y., Mitchell, M., Hughes, S., Wolf, T., Bahdanau, D., von Werra,
  L., and de~Vries, H.}
\newblock The stack: 3 tb of permissively licensed source code.
\newblock {\em Preprint\/} (2022).

\bibitem{kocher2018spectre}
{\sc Kocher, P.}
\newblock Spectre mitigations in microsoft’s c/c++ compiler.
\newblock Retrieved July 27, 2023 from
  \url{https://www.paulkocher.com/doc/MicrosoftCompilerSpectreMitigation.html},
  2018.

\bibitem{kocher2019spectre}
{\sc Kocher, P., Horn, J., Fogh, A., Genkin, D., Gruss, D., Haas, W., Hamburg,
  M., Lipp, M., Mangard, S., Prescher, T., et~al.}
\newblock Spectre attacks: Exploiting speculative execution.
\newblock In {\em 2019 IEEE Symposium on Security and Privacy (SP)\/} (2019),
  IEEE, pp.~1--19.

\bibitem{kocher1996timing}
{\sc Kocher, P.~C.}
\newblock Timing attacks on implementations of diffie-hellman, rsa, dss, and
  other systems.
\newblock In {\em Advances in Cryptology—CRYPTO’96: 16th Annual
  International Cryptology Conference Santa Barbara, California, USA August
  18--22, 1996 Proceedings 16\/} (1996), Springer, pp.~104--113.

\bibitem{agl2013ctgrind}
{\sc Langley, A.}
\newblock ctgrind: Checking that functions are constant time with valgrind.
\newblock \url{https://github.com/agl/ctgrind}, 2013.
\newblock Available: https://github.com/agl/ctgrind.

\bibitem{Lipp2018meltdown}
{\sc Lipp, M., Schwarz, M., Gruss, D., Prescher, T., Haas, W., Fogh, A., Horn,
  J., Mangard, S., Kocher, P., Genkin, D., Yarom, Y., and Hamburg, M.}
\newblock Meltdown: Reading kernel memory from user space.
\newblock In {\em 27th {USENIX} Security Symposium ({USENIX} Security 18)\/}
  (2018).

\bibitem{beatgammit2023base64js}
{\sc Little, J.}
\newblock base64-js.
\newblock \url{https://github.com/beatgammit/base64-js}, 2023.
\newblock Accessed: 2023-07-19.

\bibitem{liu2019roberta}
{\sc Liu, Y., Ott, M., Goyal, N., Du, J., Joshi, M., Chen, D., Levy, O., Lewis,
  M., Zettlemoyer, L., and Stoyanov, V.}
\newblock Roberta: A robustly optimized bert pretraining approach, 2019.

\bibitem{moore2023aesjs}
{\sc Moore, R.}
\newblock aes-js.
\newblock \url{https://github.com/ricmoo/aes-js}, 2023.
\newblock Accessed: 2023-07-19.

\bibitem{MosierAxiomatic22}
{\sc Mosier, N., Lachnitt, H., Nemati, H., and Trippel, C.}
\newblock Axiomatic hardware-software contracts for security.
\newblock In {\em Proceedings of the 49th Annual International Symposium on
  Computer Architecture\/} (New York, NY, USA, 2022), ISCA '22, Association for
  Computing Machinery, p.~72–86.

\bibitem{OleksenkoSpecFuzz18}
{\sc Oleksenko, O., Trach, B., Silberstein, M., and Fetzer, C.}
\newblock Specfuzz: Bringing spectre-type vulnerabilities to the surface.
\newblock In {\em Proceedings of the 29th USENIX Conference on Security
  Symposium\/} (USA, 2020), SEC'20, USENIX Association.

\bibitem{openai2023gpt4}
{\sc OpenAI}.
\newblock Gpt-4 technical report, 2023.

\bibitem{msvcspectre}
{\sc Pardoe, A.}
\newblock Spectre mitigations in msvc, Jan. 2018.

\bibitem{pearce2023examining}
{\sc Pearce, H., Tan, B., Ahmad, B., Karri, R., and Dolan-Gavitt, B.}
\newblock {Examining Zero-Shot Vulnerability Repair with Large Language
  Models}.
\newblock In {\em 2023 {IEEE} {Symposium} on {Security} and {Privacy} ({SP})\/}
  (2023), IEEE.

\bibitem{rodrigues2016flowtracker}
{\sc Rodrigues, B., Quint{\~a}o~Pereira, F.~M., and Aranha, D.~F.}
\newblock Sparse representation of implicit flows with applications to
  side-channel detection.
\newblock In {\em Proceedings of the 25th International Conference on Compiler
  Construction\/} (2016), pp.~110--120.

\bibitem{Schwarz2019ZombieLoad}
{\sc Schwarz, M., Lipp, M., Moghimi, D., Van~Bulck, J., Stecklina, J.,
  Prescher, T., and Gruss, D.}
\newblock {ZombieLoad}: Cross-privilege-boundary data sampling.
\newblock In {\em CCS\/} (2019).

\bibitem{tarlow2020learning}
{\sc Tarlow, D., Moitra, S., Rice, A., Chen, Z., Manzagol, P.-A., Sutton, C.,
  and Aftandilian, E.}
\newblock Learning to fix build errors with graph2diff neural networks.
\newblock In {\em Proceedings of the IEEE/ACM 42nd international conference on
  software engineering workshops\/} (2020), pp.~19--20.

\bibitem{touvron2023LLaMA}
{\sc Touvron, H., Lavril, T., Izacard, G., Martinet, X., Lachaux, M.-A.,
  Lacroix, T., Rozière, B., Goyal, N., Hambro, E., Azhar, F., Rodriguez, A.,
  Joulin, A., Grave, E., and Lample, G.}
\newblock Llama: Open and efficient foundation language models, 2023.

\bibitem{touvron2023LLaMAtwo}
{\sc Touvron, H., Martin, L., Stone, K., Albert, P., Almahairi, A., Babaei, Y.,
  Bashlykov, N., Batra, S., Bhargava, P., Bhosale, S., Bikel, D., Blecher, L.,
  Ferrer, C.~C., Chen, M., Cucurull, G., Esiobu, D., Fernandes, J., Fu, J., Fu,
  W., Fuller, B., Gao, C., Goswami, V., Goyal, N., Hartshorn, A., Hosseini, S.,
  Hou, R., Inan, H., Kardas, M., Kerkez, V., Khabsa, M., Kloumann, I., Korenev,
  A., Koura, P.~S., Lachaux, M.-A., Lavril, T., Lee, J., Liskovich, D., Lu, Y.,
  Mao, Y., Martinet, X., Mihaylov, T., Mishra, P., Molybog, I., Nie, Y.,
  Poulton, A., Reizenstein, J., Rungta, R., Saladi, K., Schelten, A., Silva,
  R., Smith, E.~M., Subramanian, R., Tan, X.~E., Tang, B., Taylor, R.,
  Williams, A., Kuan, J.~X., Xu, P., Yan, Z., Zarov, I., Zhang, Y., Fan, A.,
  Kambadur, M., Narang, S., Rodriguez, A., Stojnic, R., Edunov, S., and
  Scialom, T.}
\newblock Llama 2: Open foundation and fine-tuned chat models, 2023.

\bibitem{ridl}
{\sc van Schaik, S., Milburn, A., Österlund, S., Frigo, P., Maisuradze, G.,
  Razavi, K., Bos, H., and Giuffrida, C.}
\newblock {RIDL}: Rogue in-flight data load.
\newblock In {\em S\&{P}\/} (May 2019).

\bibitem{wangKLEESpectre2020}
{\sc Wang, G., Chattopadhyay, S., Biswas, A.~K., Mitra, T., and Roychoudhury,
  A.}
\newblock Kleespectre: Detecting information leakage through speculative cache
  attacks via symbolic execution.
\newblock {\em ACM Trans. Softw. Eng. Methodol. 29}, 3 (jun 2020).

\bibitem{wang2019oo7}
{\sc Wang, G., Chattopadhyay, S., Gotovchits, I., Mitra, T., and Roychoudhury,
  A.}
\newblock oo7: Low-overhead defense against spectre attacks via program
  analysis.
\newblock {\em IEEE Transactions on Software Engineering\/} (2019).

\bibitem{wang2017cached}
{\sc Wang, S., Wang, P., Liu, X., Zhang, D., and Wu, D.}
\newblock $\{$CacheD$\}$: Identifying $\{$Cache-Based$\}$ timing channels in
  production software.
\newblock In {\em 26th USENIX security symposium (USENIX security 17)\/}
  (2017), pp.~235--252.

\bibitem{wei2022chain}
{\sc Wei, J., Wang, X., Schuurmans, D., Bosma, M., Xia, F., Chi, E., Le, Q.~V.,
  Zhou, D., et~al.}
\newblock Chain-of-thought prompting elicits reasoning in large language
  models.
\newblock {\em Advances in Neural Information Processing Systems 35\/} (2022),
  24824--24837.

\bibitem{weiser2018data}
{\sc Weiser, S., Zankl, A., Spreitzer, R., Miller, K., Mangard, S., and Sigl,
  G.}
\newblock $\{$DATA$\}$--differential address trace analysis: Finding
  address-based $\{$Side-Channels$\}$ in binaries.
\newblock In {\em 27th USENIX Security Symposium (USENIX Security 18)\/}
  (2018), pp.~603--620.

\bibitem{jan2018microwalk}
{\sc Wichelmann, J., Moghimi, A., Eisenbarth, T., and Sunar, B.}
\newblock Microwalk: A framework for finding side channels in binaries.
\newblock In {\em Proceedings of the 34th Annual Computer Security Applications
  Conference\/} (New York, NY, USA, 2018), ACSAC '18, Association for Computing
  Machinery, p.~161–173.

\bibitem{jan2022microwalk}
{\sc Wichelmann, J., Sieck, F., P{\"a}tschke, A., and Eisenbarth, T.}
\newblock Microwalk-ci: practical side-channel analysis for javascript
  applications.
\newblock In {\em Proceedings of the 2022 ACM SIGSAC Conference on Computer and
  Communications Security\/} (2022), pp.~2915--2929.

\bibitem{wu2018eliminating}
{\sc Wu, M., Guo, S., Schaumont, P., and Wang, C.}
\newblock Eliminating timing side-channel leaks using program repair.
\newblock In {\em Proceedings of the 27th ACM SIGSOFT International Symposium
  on Software Testing and Analysis\/} (2018), pp.~15--26.

\bibitem{wu2022promptchainer}
{\sc Wu, T., Jiang, E., Donsbach, A., Gray, J., Molina, A., Terry, M., and Cai,
  C.~J.}
\newblock Promptchainer: Chaining large language model prompts through visual
  programming.
\newblock In {\em CHI Conference on Human Factors in Computing Systems Extended
  Abstracts\/} (2022), pp.~1--10.

\bibitem{wu2022ai}
{\sc Wu, T., Terry, M., and Cai, C.~J.}
\newblock Ai chains: Transparent and controllable human-ai interaction by
  chaining large language model prompts.
\newblock In {\em Proceedings of the 2022 CHI Conference on Human Factors in
  Computing Systems\/} (2022), pp.~1--22.

\bibitem{wu2023effective}
{\sc Wu, Y., Jiang, N., Pham, H.~V., Lutellier, T., Davis, J., Tan, L., Babkin,
  P., and Shah, S.}
\newblock How effective are neural networks for fixing security
  vulnerabilities.
\newblock {\em arXiv preprint arXiv:2305.18607\/} (2023).

\bibitem{yasunaga2021break}
{\sc Yasunaga, M., and Liang, P.}
\newblock Break-it-fix-it: Unsupervised learning for program repair.
\newblock In {\em International Conference on Machine Learning\/} (2021), PMLR,
  pp.~11941--11952.

\bibitem{zhang2022ultimate}
{\sc Zhang, Z., Barthe, G., Chuengsatiansup, C., Schwabe, P., and Yarom, Y.}
\newblock Ultimate slh: Taking speculative load hardening to the next level.
\newblock {\em Cryptology ePrint Archive\/} (2022).

\end{thebibliography}
\begin{appendices} 
\section{Microbenchmark of leaky functions compiled from literature}\label{sec:leaky_functions}


\begin{figure}[h]
\footnotesize
\begin{lstlisting}[style=CStyle2,linewidth=0.981\columnwidth,xleftmargin=0.35cm]
// taken from Pitchfork, Cauligi, et al.
int memory_leakage_case_1(int x, int y, int option) {
    volatile int z[3] = { 0, 2, 300 };
    z[2] = y;
    if (option > 3) {
        return z[1];
    } else {
        return z[x % 3];
    }
}
// table lookup - from DATA - Weiser et al.
unsigned char LUT[16]={0x52, 0x19, 0x3E, 0x7F,
                    0x0C, 0x5A, 0x6D, 0x2B,
                    0x3F, 0x1A, 0x7E, 0x53, 
                    0x6C, 0x5B, 0x0D, 0x37};
int memory_leakage_case_2_transform(int kval) { return LUT[kval % 16]; }
int memory_leakage_case_2(int key){
    int val = memory_leakage_case_2_transform(0);
    val+=memory_leakage_case_2_transform(key);
    return val;
}
// from CacheD paper- Wang et al
int memory_leakage_case_3(int secret){
    int table[128] = {0};
    for (int i=0; i<128; i++){
        table[i] = i;
    }
    int i, t;
    int index = 0;
    for (i=0; i<200; i++){
        index = (index+secret) % 128;
        t = table[index];
        t = table[(index) % 79];
    }
    return t;
}
const uint8_t book[10] __attribute__((aligned(64))) = { 52, 48, 55, 51, 56, 54, 50, 49, 57, 53 };
uint8_t* memory_leakage_case_4(uint8_t* msg, unsigned len) {
for (unsigned i = 0; i < len; ++i)
    msg[i] = book[msg[i]-48];

return msg;
}

\end{lstlisting}
\vspace{-0.7cm}
\label{snippet}
\end{figure}

\begin{figure}
\footnotesize
\begin{lstlisting}[style=CStyle2,linewidth=0.981\columnwidth,xleftmargin=0.35cm]
// getelement-taken from CacheAudit, Doychev et al
unsigned int A[16] = {0, 1, 2, 3, 4, 5, 6, 7,
                      8, 9, 10, 11, 12, 13, 14, 15};
int memory_leakage_case_5(int secret) {
    if (secret < 16)
        return A[secret];
}
// isDiffVul1 - taken from FlowTracker https://dl.acm.org/doi/pdf/10.1145/2892208.2892230
int branch_leakage_case_1(char *pw, char *in) {
 int i;
 for (i=0; i<16; i++) {
    if (pw[i]!=in[i]) {
        return 0;
    }
 }
 return 1;
}

// InsertionSort-taken from CacheAudit, Doychev, et al.
uint8_t * branch_leakage_case_2(uint8_t *a, int array_size){
    int i, j, index;
    for (i = 1; i < array_size; ++i){
        index = a[i];
        for (j = i; j > 0 && a[j-1] > index; j--)
            a[j] = a[j-1];
        a[j] = index;
    }
return a;
}

// eq - Time variant - taken from FlowTracker https://dl.acm.org/doi/pdf/10.1145/2892208.2892230
int branch_leakage_case_3(char *p, char *q) {
 if (p[0] != q[0])
    return false;
 else if (p[1] != q[1])
    return false;
 else
    return p[2] == q[2];
} 
// example 1-from Blazer, Antonopoulos, et al.
int branch_leakage_case_4(int high, uint low) {
    int i;
    if (high == 0) {
        i = 0;
        while(i < low) i++;
    }
    else {
        i = low;
        while(i > 0) i--;
    }
    return i;
}
// example 2-from Blazer, Antonopoulos, et al.
int branch_leakage_case_5(int high, int low) {
    int i;
    if (low > 0) { // O(2*low)
        i = 0;
        while(i<low) i++;
        while(i>0) i--;
    } else { // O(1)
        if (high == 0) { i = 5; } 
        else { i = 0; i++; }
    }
    return i;
}

\end{lstlisting}
\vspace{-0.7cm}
\label{snippet}
\end{figure}

\begin{figure}
\footnotesize
\begin{lstlisting}[style=CStyle2,linewidth=0.981\columnwidth,xleftmargin=0.35cm]
// taken from https://github.com/PLSysSec/haybale-pitchfork
int branch_leakage_case_6(int x) {
    if (x > 10) {
        return x % 200 * 3;
    } else {
        return x + 10;
    }
}
// taken from https://github.com/PLSysSec/haybale-pitchfork
int branch_leakage_case_7(int x, int y, int option) {
    volatile int z[3] = { 0, 2, 300 };
    z[2] = y;
    if (option > 3) {
        return z[1];
    } else {
        return z[2];
    }
}
// from ctgrind tool  github repo
char branch_leakage_case_8(unsigned char *a, unsigned char *b) {
  unsigned i;
  for (i = 0; i < 16; i++) {
    if (a[i] != b[i])
      return 0;
  }
  return 1;
}
// mu - taken from SC-Eliminator https://dl.acm.org/doi/pdf/10.1145/3213846.3213851
// the C code of a textbook implementation of a 3-way cipher.
int32_t * branch_leakage_case_9(int32_t *a) {
    int i;
    int32_t b[3];
    b[0] = b[1] = b[2] = 0;
    for (i=0; i<32; i++) {
        b[0] <<= 1; 
        b[1] <<= 1; 
        b[2] <<= 1;
        if(a[0]&1) 
            b[2] |= 1;
        if(a[1]&1) 
            b[1] |= 1;
        if(a[2]&1) 
            b[0] |= 1;
        a[0] >>= 1; 
        a[1] >>= 1; 
        a[2] >>= 1;
    }
    a[0] = b[0]; 
    a[1] = b[1]; 
    a[2] = b[2];

return a;
}

\end{lstlisting}
\vspace{-0.7cm}
\label{snippet}
\end{figure}

\begin{figure}
\footnotesize
\begin{lstlisting}[style=CStyle2,linewidth=0.981\columnwidth,xleftmargin=0.35cm]
// taken from https://github.com/PLSysSec/haybale-pitchfork
uint8_t branch_leakage_case_10(uint8_t* public_arr, uint8_t public_arr_len, uint8_t* secret_arr, uint8_t i) {
    uint8_t x = public_arr[i];
    for (int j = 0; j < public_arr_len; j++) {
        secret_arr[j] += x;
    }
    if (x > 10) {
        return public_arr[0] + secret_arr[0];
    } else {
        return public_arr[1] + secret_arr[1];
    }
}
// bubblesort- taken from CacheAudit https://www.usenix.org/system/files/conference/usenixsecurity13/sec13-paper_doychev.pdf
uint8_t * branch_leakage_case_11(uint8_t *a, int n){
    int i, j, temp;
    for (i = 0; i < n - 1; ++i)
        for (j = 0; j < n - 1 - i; ++j)
            if (a[j] > a[j+1]){
                temp = a[j+1];
                a[j+1] = a[j];
                a[j] = temp;
            }
    return a;
 }
// SelectionSort - taken from CacheAudit https://www.usenix.org/system/files/conference/usenixsecurity13/sec13-paper_doychev.pdf
uint8_t * branch_leakage_case_12(uint8_t *a, int array_size){
    int i;
    for (i = 0; i < array_size - 1; ++i){
        int j, min, temp;
        min = i;
        for (j = i+1; j < array_size; ++j){
            if (a[j] < a[min])
                min = j;
        }
        temp = a[i];
        a[i] = a[min];
        a[min] = temp;
    }
    return a;
}
\end{lstlisting}
\vspace{-0.7cm}
\label{snippet}
\end{figure}

\newpage

\end{appendices}
\end{document}